\documentclass[11pt]{article}
\usepackage{amsmath}
\usepackage{pstricks}
\usepackage{youngtab}
\usepackage{bbm}

\def\half{\frac{1}{2}}
\def\bsh{\backslash}

\newfont{\bbbold}{msbm10 scaled \magstep1}

\def\bbR{\mbox{\bbbold R}}

\def\bbZ{\mbox{\bbbold Z}}
\def\cA{{\cal A}}

\def\cD{{\cal D}}
\def\cE{{\cal E}}
\def\cF{{\cal F}}

\def\cH{{\cal H}}

\def\cL{{\cal L}}

\def\cN{{\cal N}}
\def\cO{{\cal O}}

\def\cR{{\cal R}}

\def\cV{{\cal V}}

\newfont{\goth}{eufm10 scaled \magstep1}

\def\gg{\mbox{\goth g}}

\def\gl{\mbox{\goth l}}

\def\go{\mbox{\goth o}}
\def\gp{\mbox{\goth p}}

\def\gs{\mbox{\goth s}}

\def\a{\alpha}
\def\b{\beta}
\def\c{\gamma}\def\C{\Gamma}
\def\d{\delta}
\def\e{\epsilon}\def\ve{\varepsilon}\def\vare{\varepsilon}
\def\f{\phi}\def\F{\Phi}
\def\h{\eta}

\def\l{\lambda}\def\L{\Lambda}

\def\S{\Sigma}

\def\th{\theta}

\def\be{\begin{equation}}\def\ee{\end{equation}}
\def\bea{\begin{eqnarray}}\def\eea{\end{eqnarray}}
\def\barr{\begin{array}}\def\earr{\end{array}}

\def\o{\omega}\def\O{\Omega}
\def\del{\partial}
\def\ua{\underline{\alpha}}
\def\ub{\underline{\phantom{\alpha}}\!\!\!\beta}
\def\uc{\underline{\phantom{\alpha}}\!\!\!\gamma}

\def\um{\underline{\mu}}

\def\xz{\times}

\def\nm{\nonumber \\}

\let\la=\label

\def\nn{\nonumber}
\def\bd{\begin{document}}
\def\ed{\end{document}}
\def\ba{\begin{array}}
\def\ea{\end{array}}
\def\bea{\begin{eqnarray}}
\def\eea{\end{eqnarray}}
\def\ft#1#2{\tfrac{#1}{#2}}
\def\fft#1#2{\frac{#1}{#2}}
\def\sst#1{{\scriptscriptstyle #1}}
\def\oneone{\rlap 1\mkern4mu{\rm l}}

\newcommand{\eq}[1]{(\ref{#1})}
\newcommand{\w}[1]{\\[0.#1cm]}
\def\eqs#1#2{(\ref{#1}-\ref{#2})}
\def\det{{\rm det\,}}
\def\tr{{\rm tr}}
\def\ad{{\rm ad}}

\newcommand{\hoch}[1]{$\, ^{#1}$}
\newcommand{\imperial}{\it\small Theoretical Physics Group, Imperial College London\\ Prince Consort Road, London SW7 2AZ, UK}
\newcommand{\kings}
{\it\small Department of Mathematics, King's College, University of London\\ Strand, London WC2R 2LS, UK}
\newcommand{\uu}
{\it\small Department of Theoretical Physics, Uppsala, Sweden}
\newcommand{\hip}
{\it\small HIP-Helsinki Institute of Physics, P.O. Box 64 FIN-00014
University of Helsinki, Suomi-Finland}
\newcommand{\stock}
{\it\small Department of Theoretical Physics, Stockholm, Sweden}
\newcommand{\golm}
{\it\small AEI, Max Planck Institut f\"ur Gravitationsphysik\\ Am M\"{u}hlenberg 1, D-14476 Potsdam, Germany}
\makeatletter
\renewcommand\theequation{\thesection.\arabic{equation}}
\@addtoreset{equation}{section} \makeatother

\newcommand{\sa}{/ \hspace{-1.2ex}}
\newcommand{\saa}{/ \hspace{-1.4ex}}
\newcommand{\saaa}{\, / \hspace{-1.6ex}}
\newcommand{\Scal}[1]{\Bigl ({#1} \Bigr )}
\newcommand{\scal}[1]{\bigl ({#1} \bigr )}

\newcommand{\CR}{\nonumber \\*}

\newcommand{\trace}{\hbox {tr}~}
\newcommand{\traceS}{\hbox {tr}_{\scriptscriptstyle \mathfrak{S}}~}

\DeclareMathAlphabet{\mathpzc}{OT1}{pzc}{m}{it}
\def\BRST{\,\mathpzc{s}\,}
\def\aBRST{{\scriptstyle (\mathpzc{s})}}
\def\q{{{\scriptscriptstyle (Q)}}}
\def\qs{{\scriptscriptstyle (Q\mathpzc{s})}}
\def\Qsla{{\mathcal{S}_{\q}}}
\def\Slav{{\mathcal{S}_\aBRST}}
\def\epsilonb{{\overline{\epsilon}}}
\def\bulletup{{\scriptstyle \bullet}}

\newcommand{\gra}[2]{{\scriptscriptstyle (#1 , #2 )}}
\newcommand{\ord}[1]{{\scriptscriptstyle (#1)}}

\def\cL{{\cal L}}
\def\cN{\mathcal{N}}
\def\cO{\mathcal{O}}

\def\ie{{\it i.e.}\ }
\def\eg{{\it e.g.}\ }

\newcommand{\sfrac}[2]{{\scriptstyle \frac{#1}{#2}}}
\newcommand{\stfrac}[2]{{\scriptscriptstyle \frac{#1}{#2}}}

 \def\balpha{{\overline{\alpha}}}
 \def\bbeta{{\overline{\beta}}}
 \def\bgamma{{\overline{\gamma}}}
 \def\bdelta{{\overline{\delta}}}
 \def\bepsilon{{\overline{\epsilon}}}
 \def\bvarepsilon{{\overline{\varepsilon}}}
 \def\bzeta{{\overline{\zeta}}}
 \def\bareta{{\overline{\eta}}}
 \def\btheta{{\overline{\theta}}}
 \def\bvartheta{{\overline{\vartheta}}}
 \def\biota{{\overline{\iota}}}
 \def\bkappa{{\overline{\kappa}}}
 \def\blambda{{\overline{\lambda}}}
 \def\bmu{{\overline{\mu}}}
 \def\bnu{{\overline{\nu}}}
 \def\bxi{{\overline{\xi}}}
 \def\bpi{{\overline{\pi}}}
 \def\brho{{\overline{\rho}}}
 \def\bvarrho{{\overline{\varrho}}}
 \def\bsigma{{\overline{\sigma}}}
 \def\bvarsigma{{\overline{\varsigma}}}
 \def\btau{{\overline{\tau}}}
 \def\bphi{{\overline{\phi}}}
 \def\bvarphi{{\overline{\varphi}}}
 \def\bchi{{\overline{\chi}}}
 \def\bpsi{{\overline{\psi}}}
 \def\bomega{{\overline{\omega}}}

\def\thalf{{\textrm{\tiny\textonehalf}}}
\def\tquarter{{\textrm{\tiny\textonequarter}}}
\def\Ko{{\scriptscriptstyle K}}
\def\tKo{\scriptscriptstyle k }
\def\corr{$\clubsuit$}

\newcommand{\auth}{\large J. Greitz${}^{a,b,}$\footnote{email:jesper.greitz@nordita.org}, P.S.\ Howe${}^{a,}$\footnote{email: paul.howe@kcl.ac.uk}}

\begin{document}

\renewcommand{\thefootnote}{\fnsymbol{footnote}}

\null
\begin{flushright}
{\small KCL-MTH-12-02}\\
{\small NORDITA-2012-20}
\vskip 1.5 cm
\end{flushright}

\begin{center}
{\Large{\bf Half-maximal supergravity in three dimensions: supergeometry, differential forms and algebraic structure}}
\vspace{.75cm}

\auth
\end{center}
\vspace{.5cm}

\centerline{${}^a${\it \small Department of Mathematics, King's College London}}
\centerline{{\it \small The Strand, London WC2R 2LS, UK}}
\vspace{.5cm}
\centerline{${}^b${\it \small Nordita}}
\centerline{{\it \small Royal Institute of Technology and Stockholm University }}
\centerline{{\it \small Roslagstullsbacken 23, SE-106 91 Stockholm, Sweden }}

\vspace{1cm}

\centerline{{\bf Abstract}}
\vskip .5cm
The half-maximal supergravity theories in three dimensions, which have local $SO(8)\xz SO(n)$ and rigid $SO(8,n)$ symmetries, are discussed in a superspace setting starting from the superconformal theory. The on-shell theory is obtained by imposing further constraints; it is essentially a non-linear sigma model that induces a Poincar\'e supergeometry. The deformations of the geometry due to gauging are briefly discussed. The possible $p$-form field strengths are studied using supersymmetry and $SO(8,n)$ symmetry. The set of such forms obeying consistent Bianchi identities constitutes a Lie super co-algebra while the demand that these identities admit solutions places a further constraint on the possible representations of $SO(8,n)$ that the forms transform under which can be easily understood using superspace cohomology. The dual Lie superalgebra can then be identified as the positive sector of a Borcherds superalgebra that extends the Lie algebra of the duality group. In addition to the known
$p=2,3,4$ forms, which we construct explicitly, there are five-forms that can be non-zero in supergravity, while all forms with $p>5$ vanish. It is shown that some six-forms can have non-trivial contributions at order $\a'$. 

\vspace{1cm}


\renewcommand{\thefootnote}{\arabic{footnote}}
\setcounter{footnote}{0}

\pagebreak
\tableofcontents
\setcounter{page}{1}


\section{Introduction}


Rigid symmetry groups play an important r\^ole in maximal supergravity theories in various dimensions \cite{Cremmer:1979up}. In particular, in $D=3$, this group is $E_8$ \cite{Marcus:1983hb}, the largest finite such group. The scalars in this theory live in the coset $SO(16)\bsh E_8$, where the R-symmetry group $SO(16)$ is local, and in fact the theory is essentially a sigma model because there are no purely gravitational degrees of freedom. In superspace this sigma model induces a super-geometry which was described in some detail in \cite{Greitz:2011vh}. The rigid symmetry group plays a crucial r\^ole in the construction of the additional field strength form fields and also facilitates the gauging of the theory. In addition these forms are associated with extended algebraic structures in maximal supergravity theories including Borcherds algebras \cite{HenryLabordere:2002dk,HenryLabordere:2002xh} and $E_{11}$ \cite{Julia:1997cy,West:2001as,Riccioni:2007au}. 

There is also an interesting set of theories that have half-maximal supersymmetry.  In $D=3$ there are a number of half-maximal supergravities with sigma models of the form $(SO(8)\xz SO(n))\bsh SO(8,n)$ that were first introduced in \cite{Marcus:1983hb} and further studied in \cite{de Wit:1992up}. The forms in these models have been discussed in \cite{Bergshoeff:2007vb} while the gauged versions were studied  in \cite{Nicolai:2001ac,deWit:2003ja}.

In this paper we study these half-maximal theories in a superspace setting, starting, as in the maximal case, from the off-shell superconformal geometry. We then describe how one needs to specify the fields that appear in this geometry in terms of the physical sigma model fields in order to obtain the on-shell Poincar\'e supergravity theory. An interesting feature of this geometry is that the local symmetry group is bigger that the R-symmetry group $SO(8)$ because the  $SO(n)$ curvature has to be included. We also study the gauging in this geometrical setting. A feature here is that, although the off-shell superconformal geometry is special in $N=8$ because one can impose a duality constraint on the dimension-one scalars, it turns out that this is not sufficient in the gauged case and that one needs to keep both dualities. 

We then go on to study the form fields in these theories using only supersymmetry and the bosonic symmetries of the Poincar\'e theory. Assuming that the forms fall into representations of $SO(8,n)$ and demanding that the Bianchi identities be consistent we can classify the possible forms that can arise. For potentials up to the spacetime limit, i.e. field strength $p$-forms with $p=2,3,4$, we find agreement with the results of \cite{Bergshoeff:2007vb}, as one would expect. We then find all the allowed five-forms. Such forms can have non-vanishing dimension-zero components in supergravity and so need to be classified in order to find the complete theory. They can also play a r\^ole in the gauge hierarchy \cite{deWit:2008gc} although we shall not discuss this topic in detail here. The interested reader can find a superspace discussion for the maximal case in \cite{Greitz:2011vh}.  We do not solve all of the Bianchi identities for the forms but we do give some examples of solutions. 

The allowed degrees of the forms are naturally truncated in spacetime, but can increase without limit in superspace because the odd basis forms commute. Form fields beyond the spacetime limit were discussed for maximal theories in $D=10$ in \cite{Greitz:2011da}. There it was shown that the degrees of the forms can indeed be increased without limit and that these forms transform under the representations that one would expect from the Borcherds algebra point of view \cite{HenryLabordere:2002dk,Slansky:1991dx}. In a sense this is quite satisfying because to get the full Borcherds algebra one needs all of the forms, whereas the spacetime approach inevitably leads to a truncated picture. More generally, one can show that the set of field-strength forms that satisfy consistent Bianchi identities gives rise to a Lie super co-algebra. The forms transform under representations of the duality group (here $SO(8,n)$) and there is a further constraint on the allowed representations that arises from the requirement that the Bianchi identities be not only consistent but also allow solutions. In the half-maximal $D=3$ case it turns out, for cohomological reasons, that there is only one such constraint; it restricts the allowed three-form field strengths. All the higher-degree forms are then determined by the consistency of the Bianchi identities. This final constraint restricts the algebraic structure and one can show that the dual Lie superalgebra can be identified with the positive sector of a Borcherds algebra, i.e. it is spanned by the elements of the algebra that correspond to the positive roots. This set of roots is generated by the positive simple roots of $\gs\go(8,n)$ together with an extra odd root.

The field-strength forms with degree greater that $D+2$ are trivially zero in the supergravity limit, so that one would like to see if any of them could become non-zero if one includes string corrections. In maximal supergravity  such corrections start at $\a'^3$ and are consequently not easy to analyse. On the other hand, in the half-maximal $D=3$ theories, one might expect there to be corrections starting at order $\a'$. We investigate this possibility here by looking at a subset of the possible six-forms that can arise and give some evidence that one can indeed find some non-vanishing six-form components that are compatible with at least some of the Bianchi identities. This result gives us confidence that forms beyond the spacetime limit are indeed physically significant when one takes higher-order corrections into account.

The organisation of the paper is as follows: in section 2 we describe the geometrical set-up and review the off-shell superconformal constraints for $N$-extended supergravity in $D=3$. In section 3 we introduce the $SO(8,n)$ sigma model in the context of this supergravity background and show how the latter can accommodate it by making appropriate identifications. We also introduce the vector fields that transform under the adjoint representation of $SO(8,n)$. We briefly discuss the deformation of the geometry due to gauging via the modified Maurer-Cartan equation and show that both dualities of the dimension-one scalar superfield are required for this to work. In section 4  we turn our attention to the additional form fields. As well as the duals to the scalars (two-form field strengths) there are three-, four- and five-forms whose potentials have no physical degrees of freedom, and higher-degree forms whose field strengths are identically zero in supergravity. We discuss the Lie super co-algebra associated with these forms and argue that the dual Lie superalgebra is the positive sector of a Borcherds algebra. In section 5 we identify some possible six-forms in the theory and show that some of them can be non-zero when order $\a'$ corrections are switched on. Our conclusions are given in section 6.


\section{Geometry}


\subsection{Conformal constraints}


For $N$-extended supersymmetry we consider a supermanifold $M$ with (even$|$odd)-dimension $(3|2N)$. The basic structure is determined by a choice of odd tangent bundle $T_1$ such that the Frobenius tensor, which maps pairs of sections of $T_1$ to the even tangent bundle, $T_0$, generates the latter. We shall also suppose that there is a preferred basis $E_{\a i},\,\a=1,2;\,i=1,\ldots N$ for $T_1$ such that the components of the Frobenius tensor, which we shall also refer to as the dimension-zero torsion, are

\be \label{conv. constr.}
 T_{\a i\b j}{}^c=-i\d_{ij} (\c^c)_{\a\b}\,;\qquad c=0,1,2\ .
 \la{2.1}
\ee

At this stage $T_0$ is defined as the quotient, $T/T_1$, but we can make a definite choice for $T_0$ by imposing some suitable dimension one-half constraint. When this has been done, the structure group will be reduced to $SL(2,\bbR)\xz SO(N)$, with the Lorentz vector indices being acted on by the local  $SO(1,2)$ associated with $SL(2,\bbR)$. The dimension-zero torsion \eq{2.1} is also invariant under local Weyl rescalings, although we shall not include this factor in the structure group. This indicates that we can expect to find a conformal multiplet.  With respect to this structure we have preferred basis vector fields $E_A=(E_a,E_{\ua})=(E_a,E_{\a i})$ with dual one-forms $E^A=(E^a,E^{\ua})=(E^a, E^{\a i})$, the latter being related to the coordinate basis forms $dz^M=(dx^m, d\th^{\um})$ by the supervielbein matrix $E_M{}^A$, i.e. $E^A=dz^M E_M{}^A$. Here, coordinate indices are taken from the middle of the alphabet, preferred basis indices from the beginning, while even (odd) indices are latin and greek respectively. Underlined odd indices run from 1 to $2N$, and $SO(N)$ vector indices are denoted $i,j$ etc. 

We now introduce a set of connection one-forms, $\O_A{}^B$, for the above structure group. We have

\bea
 \O_a{}^{\ub}&=& \O_{\ua}{}^b=0\nn\w1
 \O_{\a i}{}^{\b j}&=& \d_i{}^j \O_\a{}^\b + \d_\a{}^\b \O_i{}^j\nn\w1
 \O_a{}^b&=& - (\c_a{}^b)_\a{}^\b \O_\b{}^\a\ .
 \la{2.2}
\eea

Spinor indices $\a,\b$ are raised and lowered by the epsilon tensor, while Lorentz and $SO(N)$ vector indices are raised by the corresponding metrics $\h_{ab}, \d_{ij}$. We have $\O_{\a\b}=\O_{\b\a}$ while $\O_{ab}$ and $\O_{ij}$ are antisymmetric. The torsion and curvature are defined in the usual way

\bea
 T^A&=& DE^A:=d E^A + E^B \O_B{}^A\nn\w1
 R_A{}^B&=& d\O_A{}^B + \O_A{}^C \O_C{}^B\ .
 \la{2.3}
\eea

The Bianchi identities are

\bea
 DT^A&=& E^B R_B{}A\nn\w1
 D R_A{}^B&=&0\ .
 \la{2.4}
\eea

Equation \eq{2.1} does not simply determine the structure group, it is also a constraint. With an appropriate choice of dimension one-half connections and of $T_0$, and making use of the dimension one-half Bianchi identity, one finds that all components of the dimension one-half torsion may be set to zero:

\be
 T_{\ua\ub}{}^{\uc}=T_{a \ub}{}^c=0\ .
 \la{2.5}
\ee

Imposing further conventional constraints corresponding to the dimension-one connection components we find that the dimension-one torsion can be chosen to have the form

\bea
 T_{ab}{}^c&=&0\nn\w1
 T_{a \b j}{}^{\c k}&=&   (\c_a)_\b{}^\c K_j{}^k + (\c^b)_\b{}^\c L_{ab j}{}^k\ ,
 \la{2.6}
\eea

where $K_{ij}$ is symmetric and $L_{abij}$ is antisymmetric on both pairs of indices. The dimension-one curvatures are

\bea
 R_{\a i\b j, cd}&=& -2i(\c_{cd})_{\a\b} K_{ij} -2i\ve_{\a\b} L_{cdij}\nn\w1
 R_{\a i\b j,kl}&=& i\ve_{\a\b}(M_{ijkl} + 4 \d_{[i[k} K_{j]l]})-i(\c^a)_{\a\b} (4\d_{(i[k} L_{a j) l]}-\d_{ij} L_{a kl})\ ,
 \la{2.7}
\eea

where $L_{ab}=\ve_{abc} L^c$, and $M_{ijkl}$ is totally antisymmetric. This geometry  describes an off-shell superconformal multiplet \cite{Howe:1995zm}. The interpretation of the dimension-one fields, $K,L,M$, is as follows. The geometry is determined by the basic constraint \eq{2.1} which is invariant under Weyl rescalings where the parameter is an unconstrained scalar superfield. This means that some of the fields that appear in the geometry do not belong to the conformal supergravity multiplet. At dimension one $K$ and $L$ are of this type, so that we could set them to zero if we were only interested in the superconformal multiplet. The field $M_{ijkl}$, on the other hand, can be considered as the field strength superfield for the conformal supergravity multiplet \cite{Howe:1995zm}.\footnote{This was discussed explicitly in  for the case of $N=8$ in \cite{Howe:2004ib}.} The fact that $M$ is not expressible in terms of  the torsion is due to a lacuna in Dragon's theorem  \cite{Kuzenko:2011xg,Cederwall:2011pu} which in higher-dimensional spacetimes states that the curvature is so determined \cite{Dragon:1978nf}. We recall that in three-dimensional spacetime there is no Weyl tensor but that its place is taken by the dimension-three Cotton tensor. This turns out to be a component of the superfield $M_{ijkl}$ so that we could refer to the latter as the super Cotton tensor. Using the notation $[k,l]$ to denote fields that have $k$ antisymmetrised $SO(N)$ indices and $l$ symmetrised spinor indices, one can see that the component fields of the superconformal multiplet fall into two sequences starting from $M_{ijkl}$. The first has fields of the type $[4-p,p]$, where the top ($[4,0]$) component is the supersymmetric Cotton tensor, while the second has fields of the type $(4+p,p)$ and therefore includes higher spin fields for $N>8$. There is also a second scalar $[4,0]$ at dimension two. Fields with two or more spinor indices obey covariant conservation conditions so that each field in the multiplet has two degrees of freedom multiplied by the dimension of the $SO(N)$ representation, provided that we count the dimension-one and -two scalars together. It is easy to see that the number of bosonic and fermionic degrees of freedom in this multiplet match. 


\subsection{The $N=8$ case}


The case $N=8$ is special for two reasons. Firstly, it is possible to impose a self-duality constraint on the superfield $M_{ijkl}$ that reduces the size of the conformal supergravity multiplet to $128 + 128$. The fields are the graviton, 8 gravitini, the $SO(8)$ gauge fields, the dimension-one scalars $M_{ijkl}$, a matching dimension-two set with opposite duality and $56$ dimension three-halves spinor fields (three-index antisymmetric field $\l_{\a ijk}$). The second feature is that it is possible in this case to take the R-symmetry group to be $Spin(8)$ rather than $SO(8)$. It turns out that this is the correct choice in order to describe the Poincar\'e theories, and so we shall switch to this for the remainder of the paper. We denote the spinor indices by $I,J,\ldots$ ($(0010)$ representation) and $I'.J'\ldots$ ($(0001)$ representation), while we keep $i,j,\ldots$ for the vector representation $(1000)$. All three types of index can take 8 values.  So for $N=8$ we shall take the basis odd one-forms to be $E^{\a I}$, and in the above formulae replace all the small internal indices by capital ones. Thus the non-zero components of the torsion are, at dimension zero

\be
 T_{\a I\b J}{}^c=-i\d_{IJ} (\c^c)_{\a\b}\,;\qquad c=0,1,2\ ,
 \la{2.8}
\ee

and, at dimension one,

\be
 T_{a \b J}{}^{\c K}=   (\c_a)_\b{}^\c K_J{}^K + (\c^b)_\b{}^\c L_{ab J}{}^K\ .
 \la{2.9}
\ee

The dimension-one curvatures are

\bea
 R_{\a I\b J, cd}&=& -2i(\c_{cd})_{\a\b} K_{IJ} -2i\ve_{\a\b} L_{cdIJ}\nn\w1
 R_{\a I\b J,KL}&=& i\ve_{\a\b}(M_{IJKL} + 4 \d_{[I[K} K_{J]L]})-i(\c^a)_{\a\b} (4\d_{(I[K} L_{a J) L]}-\d_{IJ} L_{a KL})\ ,
 \la{2.10 }
\eea

The field $M_{IJKL}$ can be self- or anti-self-dual; in the former this is the representation $(2000)$, i.e. a symmetric traceless second-rank tensor, while in the second case the representation is $(0002)$ which is the anti-self-dual fourth-rank tensor. 

The other components of the curvature and torsion can be derived straightforwardly from here, although we shall not give the details in this paper.

For the conformal case, the dimension three-halves Bianchi identities were solved explicitly in \cite{Kuzenko:2011xg}, while a detailed discussion of the $N=8$ case has been given in \cite{Cederwall:2011pu}.


\subsection{The on-shell Poincar\'e theory}


As the graviton and the gravitino carry no physical degrees of freedom in on-shell Poincar\'e supergravity in $D=3$, it is necessary to introduce matter fields in order to get a non-trivial theory. For $N=16$ this takes the form of a supersymmetric sigma model based on the coset $SO(16)\bsh E_8$ while for $N=8$ there is a series of such models that make use of the cosets $(SO(8)\xz SO(n))\bsh SO(8,n)$. Notice this implies that the local R-symmetry group will be enlarged by the $SO(n)$ factor and hence that there will be a corresponding additional curvature tensor in the geometry. We shall denote $SO(8,n)$ vector indices by $R,S,\ldots$ and $SO(n)$ vector indices by $r,s\ldots$, so $R=(i,r)$ where $i$ is an $SO(8)$ vector index as before. We take the generators of $\gs\go(8,n)$, $M_{RS}=-M_{SR}$, to satisfy

\be
[M_{RS}, M^{TU}]= -4\d_{[R}{}^{[T} M_{S]}{}^{U]}
\la{2.11}
\ee

Written out with respect to the $\gs\go(8)\oplus\gs\go(n)$ decomposition this is

\bea
 [M_{ij}, M^{kl}]&=& -4\d_{[i}{}^{[k} M_{j]}{}^{l]}\nn\w1
 [M_{rs},M^{tu}]&=& -4\d_{[r}{}^{[t} M_{s]}{}^{u]}\nn\w1
 [M_{ij}, M^{kr}]&=&-2\d_{[i}{}^k M_{j]}{}^r\nn\w1
 [M_{rs}, M^{kt}]&=&2\d_{[r}{}^t M_{s]}{}^k\nn\w1
 [M_{ir},M_{js}]&=& - \h_{ij} M_{rs}- \h_{rs} M_{ij}\ .
 \la{2.12}
\eea

The $SO(8,n)$ metric is $\h_{RS}=(\d_{ij},-\d_{rs})$.

The sigma model field $\cV$ is an element of $SO(8,n)$ that depends on the superspace coordinates. It is acted on to the right by $SO(8,n)$ and to the left by the local $SO(8)\xz SO(n)$ and therefore corresponds to an $(SO(8)\xz SO(n))\bsh SO(8,n)$ sigma model superfield. The Maurer-Cartan form is

\be
 \F:=d\cV \cV^{-1}:=P+ Q\ ,
 \la{2.13}
\ee

where $Q=\half \O^{ij} M_{ij}+\half \O^{rs} M_{rs}$, $ \O^{rs}$ being the $\gs\go(n)$ connection and where $P=P^{ir} M_{ir}$ takes its values in the quotient algebra. From the Maurer-Cartan equation (vanishing $SO(8,n)$ curvature), $d\F+\F^2=0$, we find

\bea
 DP&=&0\w1
 R&=&-P^2\ ,
 \la{2.14}
\eea

where $R:= \half R^{ij} M_{ij}+\half R^{rs} M_{rs}$ is the $\gs\go(8)\oplus\gs\go(n)$ curvature, while $D$ is the corresponding covariant exterior derivative. In indices, the above equations are

\bea
 2 D_{[A} P_{B]} + T_{AB}{}^C P_C&=& 0\w1
 \la{3.4}
 R_{AB}&=&  [P_A, P_B]\ .
 \la{2.15}
\eea

The dimension of the sigma model coset is $8n$, so we need an equal number of fermions for supersymmetry. To ensure this we impose a constraint on the dimension one-half component of $P$. We set

\be
 P_{\a I}^{ir}=i(\S^i)_{IJ'}\L^r_{\a J'}\ ,
 \la{2.16}
\ee

where $\L^r_{\a J'}$ describes the $8n$ physical one-half fields. The dimension-one component of \eq{3.4} is then satisfied if

\be
 D_{\a I} \L^r_{\b J'}=\half(\c^a)_{\a\b} (\S_i)_{IJ'} P_a^{ir}\ .
 \la{2.17}
\ee

We can think of $P_{a I}$ as essentially the spacetime derivative of the physical scalar fields. In order to see this more explicitly, it is perhaps useful to look at the linearised limit. In the physical gauge we can put $\cV=\exp (\f^{ir} M_{ir})$ where $\f^{ir}$ denotes the $8n$ scalars. If we now keep only terms linear in the fields we find

\bea
 D_{\a I} \f^{ir}&=& i (\S^i \L^r_\a)_I \nn\w1
 D_{\a I}\L^r_{\b J'}&=& \half(\c^a)_{\a\b} (\S_i)_{IJ'} P^{ir}_a=\half(\c^a)_{\a\b} (\S_i)_{IJ'} \del_a \f^{ir}\ ,
 \la{2.18}
\eea

where $D_{\a I}$ here is now the usual supercovariant derivative in flat superspace. It follows from \eq{2.18} that, in the linearised limit, both $\f^{ir}$ and $\L^r_{\a I'}$ satisfy free field equations of motion. To see this explicitly one needs to apply another spinorial derivative to the second of these equations and use the supersymmetry algebra to find the Dirac equation. The scalar equation then follows from this by applying another derivative.

It is now easy to compute the dimension-one curvature and torsion in terms of the sigma model fields and to verify that they can be slotted into the superconformal geometry. We find

\bea
 K_{IJ}&=& \frac{i}{4}\d_{IJ}B:= \frac{i}{4}\d_{IJ}\L\L\nn\w1
 L_{a IJ}&=&\, \frac{i}{8} (\S^{ij})_{IJ} A_{a ij}:= \frac{i}{8} (\S^{ij})_{IJ} \L\c_a\S_{ij} \L\nn\w1
 M_{IJKL}&=&\frac{i}{32}(\S^{ij})_{IJ}(\S^{kl})_{KL} B_{ijkl}:= \frac{i}{32}(\S^{ij})_{IJ}(\S^{kl})_{KL} \L\S_{ijkl}\L\ ,
 \la{2.19}
\eea

where, on the right-hand-side, the spacetime and internal spinor indices are contracted in the natural way (see appendix). The internal $SO(n)$ vector indices are contracted with $\h_{rs}$. These formulae determine the non-zero dimension-one torsion and curvature components. The dimension-one component of the $SO(n)$ curvature is

\be
R_{\a I\b J,rs}=2(\c^a)_{\a\b} \d_{IJ} \L_r\c_a\L_s -\half\ve_{\a\b} (\S^{ij})_{IJ} \L_r\S_{ij}\L_s\ .
\la{2.20}
\ee

Notice that $M_{IJKL}$ is in the representation $(0002)$ (because $\L$ carries a primed spinor index), so that it is anti-self-dual. 

Equations \eq{2.19} and \eq{2.20} show that the geometry is determined in terms of the matter fields, and so the full non-linear equations of motion for the physical fields can be derived from the above set of  equations by supersymmetry. As the details of the higher-dimensional torsion and curvature components and the equations of motion are rather similar to the maximal case we shall not give them here, but refer the interested reader
to \cite{Greitz:2011vh} where this discussion is given.


\subsection{Vector fields}


In section three we shall describe the various $p$-form fields that can arise in the theory. However, in order to understand the gauged geometry we shall only need the two-form field strengths. These should transform according to a representation of the duality group $SO(8,n)$ and by Hodge duality there should be the same number of them as there are scalars. This is accomplished by taking the vector fields to transform under the adjoint representation of $SO(8,n)$. It will turn out that $8n$ of the field strengths are essentially duals of the field strengths for the scalars at dimension one while the others are composite. In the ungauged theory the Bianchi identities for the two-forms  are abelian,

\be
dF^{RS}=0\ .
\la{2.21}
\ee

It is not difficult to solve for the components of $F^{RS}$ in terms of the physical fields. We denote the components of the scalar field matrix $\cV$ in the fundamental representation by 

\be
\cV_{\bar R}{}^R=(V_i{}^R, V_r{}^R)\ .
\la{2.22}
\ee

Then the components of $F^{RS}$ are 

\bea
F^{RS}_{\a I\b J}&=&i\ve_{\a\b} (\S^{ij})_{IJ} V_i{}^R V_j{}^S \nn\w1
F^{RS}_{a \b J}&=&   -2i(\c_a\S^i \L^r)_{\b J} V_r{}^{[R} V_i{}^{S]}\nn\w1
F^{RS}_{a b}&=&\ve_{ab}{}^c( 2P_c^{ir} V_r{}^{[R} V_i{}^{S]} + \frac{3i}{4}A_a^{ij} V_i{}^R V_j{}^S -2iA_a^{rs} V_r{}^R V_s{}^S) \ .
\la{2.23}
\eea

The bilinear (in $\L$) $A_{aij}$ is defined in \eq{2.19} above, while $A_{a rs}:=\L_r\c_a\L_s$. Notice that this equation shows that the dimension-one component of $F^{RS}$ contains the  $8n$ scalar field strengths $P_a^{ir}$ as required.


\subsection{Gauging}


Supergravity theories are gauged with the aid of the embedding tensor. The $D=3$ half-maximal gauged theories were discussed in \cite{Nicolai:2001ac,deWit:2003ja}.\footnote{For gaugings of half-maximal theories in general dimensions, see, for example \cite{Weidner:2006rp}.} In $D=3$ in the maximal theory, with $E_8$ duality group, the embedding tensor is a projector in the adjoint representation \cite{Nicolai:2000sc,Nicolai:2001sv}, which in that case coincides with the fundamental. In the half-maximal case we can use a similar approach, that is, we can take the embedding tensor $\cE_X{}^Y$ to be a projector in the adjoint representation, $X=[RS]$. This matrix, when the second index is lowered, is symmetric and projects onto the Lie algebra of the gauge group $\gg_0$. There is an additional constraint that follows because $\cE$ should be invariant under gauge transformations; this is

\be
\cE_X{}^{X'} \cE_{(Y}{}^{Y'} f_{Z)X'Y'}=0\ ,
\la{2.24}
\ee

where $f_{XYZ}$ denotes the $\gs\go(8,n)$ structure constants. 

The gauged theory has a local gauge group $G_0$, embedded in $G=SO(8,n)$ as described above, and we can also use a formalism in which the local $SO(8)\xz SO(n)$ symmetry is maintained. Thus the formalism appears $G$-covariant, but in fact is not due to the presence of the embedding tensor.

The discussion is best approached via the gauged Maurer-Cartan form \cite{de Wit:1982ig} (see \cite{Howe:1981tp} for the superspace version) which can be written

\be
\F=\cD\cV \cV^{-1}=P+ Q\ ,
\la{2.25}
\ee

where $\cD$ is a gauge-covariant derivative (for $G_0$) that acts on the $E_8$ index carried by $\cV_{\bar R}{}^R$, i.e. the superscript. The gauged Maurer-Cartan equation, which follows directly from \eq{2.25}, is

\be
R+ DP+ P^2 = g\cF:=g\cV \cF\cV^{-1}\ .
\la{2.26}
\ee

Here, $g$ is a constant with dimensions of mass which characterises the deformation and  $D$ is covariant with respect to both $SO(8)\xz SO(n)$ and $G_0$. The theory has both of these groups as local symmetries, but the rigid $SO(8,n)$ is broken. The technique we shall use in the following analysis is to work with $SO(8)\xz SO(n)$ indices, so that the gauge group is hidden from view.

The original geometrical constraint in superspace \eq{2.1}, i.e. taking the dimension-zero torsion to be the same as in flat space, together with the allowed conventional constraints, leads to the dimension-one torsion and curvatures given in equations \eq{2.9} and \eq{2.10}. Since the deformation parameter $g$ has dimension one it follows that we can expect changes to the tensors $K_{IJ}, L_{a IJ}$ and $M_{IJKL}$. These can only be proportional to $g$ multiplied by functions of the scalars and so $L_{a IJ}$ must be unchanged. This leaves $K$ and $M$.

To implement the gauging explicitly we first need  to solve for the two-form field strength. This should be projected along $\gg_0$  which leads us to propose that it should have the form

\be
\cF^X=F^Y\cE_Y{}^X\ .
\la{2.27}
\ee

It is easy to see, using the fact that $\cD\cE_X{}^Y=0$, that the Bianchi identity for $\cF^X$ will be solved if we take the components of $F^X$ to have the same form as in the ungauged case. In fact, the only $g$-dependence could be at dimension one, but since this component of $F$ is a spacetime two-form this cannot arise.  At dimension one we therefore find

\be
\cF_{\a I\b J}^{RS}=i\ve_{\a\b} (\S^{ij})_{IJ} V_i^T V_j^U \cE_{TU,}{}^{RS}\ ,
\la{2.28}
\ee

where we have replaced the adjoint indices on $\cE$ by pairs of antisymmetrised vector indices. Using this and \eq{2.26} we find that the deformations of the dimension-one geometrical tensors due to gauging have the form

\bea
R_{\a I\b J,kl}(g) &=& g\ve_{\a\b} (\S^{ij})_{IJ} f_{ij,kl}\nn\w1
R_{\a I\b J,rs}(g) &=& g\ve_{\a\b} (\S^{ij})_{IJ} f_{ij,rs}\nn\w1
D_{\a I} \L_{\b J kr}(g) &=& g\ve_{\a\b} (\S^{ij})_{IJ} f_{ij,kr}\ ,
\la{2.29}
\eea

where the functions $f$ are defined by

\be
f_{\bar R\bar S,\bar T\bar U}:=V_{\bar R}{}^R V_{\bar S}{}^S V_{\bar T}{}^T V_{\bar U}{}^U \cE_{RS,TU}\ .
\la{2.30}
\ee

Since $\cE_{XY}$ is symmetric the representations that it contains are four-index antisymmetric, two-index symmetric traceless, a singlet and a tensor with the symmetries of the Weyl tensor. In Young tableaux,

{\small \be\Yvcentermath1
\left(\yng(1,1)\, \otimes\, \yng (1,1)\right)_{\rm sym}=\yng(1,1,1,1)\, \oplus \,\yng(2)\, \oplus 1 \,\oplus \yng(2,2)
\ee}

\be
\la{2.31}
\ee

If the Weyl tensor representation were non-zero, then there would be a contribution of the same symmetry type to $f_{ij,kl}$ which cannot be accommodated in $M$ or $K$. So this representation must be absent, and there is therefore an extra constraint on $\cE$. There are no problems with any of the other representations but there is an interesting point concerning the 35-dimensional representations that appear in $f_{ij,kl}$. In fact, all three can occur: the anti-self-dual four-form will deform $M_{IJKL}$ while the self-dual four-form will deform the traceless part of $K_{IJ}$. The symmetric traceless 35 in $f_{ij,kl}$, then modifies the self-dual part of $M_{IJKL}$. Thus, in the generic gauged theory, it is not possible to impose the duality constraint on the $N=8$ superconformal multiplet.

Explicitly,  we find that the deformations of the dimension-one scalar functions are given by

\bea
K_{IJ}&=&g(\d_{IJ} f_0 + (\S^{ijkl})_{IJ} f^{(+)}_{ijkl})\nn\w1
M_{IJKL}&=&g(\S^{ij})_{IJ} (\S^{kl})_{KL} (f^{(-)}_{ijkl} + \d_{ik} f_{jl})\nn\w1
D_{\a I}\L^{\a}_{J' r}&=&g ( (\S^{ijk})_{IJ'} f_{ijkr} + (\S^i)_{IJ'} f_{ir})\ ,
\la{2.32}
\eea

where the functions on the right are in the irreducible representations indicated, with the plus and minus signs standing for self-dual and anti-self-dual respectively.



\section{Forms}


\subsection{General Method}


The allowed forms in a supergravity theory are the physical forms, their duals and any others that may be generated from this set. These typically include $(D-1)$- and $D$-form potentials, but in superspace there can also be potential forms of degree greater than the dimension of spacetime. Indeed, this leads to an infinite set of forms and an algebraic structure which will be discussed shortly. 

In the half-maximal $D=3$ case the physical bosons are the scalars which means that the dual field strengths are two-forms as we have discussed previously. The other forms can be constructed by examining all the possible Bianchi identities of the form

\be
d F_{\ell+1}=\sum_{m+n=\ell} F_{m+1}\wedge F_{n+1}\ .
\la{3.1}
\ee

Here, $\ell$ denotes the degree of the corresponding potential form on the left-hand side, and all the forms appearing in \eq{3.1} transform according to (in general, reducible) representations, $\cR_{\ell}, \cR_m, \cR_n$, of the duality group $SO(8,n)$. The idea is that one starts with the two-forms and then proceeds step by step. For example, the Bianchi identities for the three-forms will have $F_2\wedge F_2$ on the right-hand side and can thus be in the representations

{\small \be\Yvcentermath1
\left(\yng(1,1)\, \otimes\, \yng (1,1)\right)_{\rm sym}=\yng(1,1,1,1)\, \oplus \,\yng(2)\, \oplus 1 \,\oplus \yng(2,2)\ .
\la{3.2}\ee}

There are two consistency requirements that constrain the possible forms. The first is that the Bianchi identities \eq{3.1} have to be consistent, so applying $d$ to the right-hand side must give zero, and the second is that they must admit solutions. For example, for the three-forms, since $d F_2=0$, it follows that all of the above representations obey the first requirement. However, it turns out that the Bianchi identity for the Weyl-tensor representation, i.e. the last one in \eq{3.2}, does not admit a solution and must therefore be discarded. We shall now give a simple cohomological argument to show that this is the only restriction of the second type in supergravity. If we rewrite \eq{3.1} in the form

\be
I_{\ell + 2}=d F_{\ell+1}-\sum_{m+n=\ell} F_{m+1}\wedge F_{n+1}\ ,
\la{3.3}
\ee

then the first consistency condition is 

\be
dI_{\ell +2}=0\qquad {\rm modulo\ lower\ degree}\  Is\ .
\la{3.4}
\ee

Since we shall be solving the Bianchi identities sequentially, it follows that at a given level $\ell$ we can assume that the lower level identities have been solved so that we can take the right-hand side of \eq{3.4} to be zero. To analyse whether or not there will be any obstruction to solving the identities, given that the lower ones admit solutions, it will be useful to write any $n$-form as a sum of $(p,q)$ forms, $p+q=n$,  where $p\,(q)$ denotes the number of even (odd) indices that the given component has. The dimension of the $(p,q)$ component of a field strength is $1-q/2$ while the dimension of the $(p,q)$ component of a Bianchi identity is $2-q/2$. In supergravity there are no fields with negative dimensions so this means that the lowest possible non-zero component (i.e. the one with the least number of odd indices) of $F_{\ell+1}$ is $F_{\ell -1,2}$ while for the Bianchi identity $I_{\ell +2}$ it will be $I_{\ell -2, 4}$. When we write out \eq{3.4} in terms of its (even,odd) components the lowest non-vanishing component will therefore read

\be
t_0 I_{\ell-2,4}=0\ ,
\la{3.5}
\ee

where $t_0$ is the component of $d$ with bi-degrees $(-1,2)$ which essentially corresponds to multiplying a given form by the dimension-zero torsion, contracting one of the even indices with the vector index of the torsion and then symmetrising over all of the odd indices. (See the appendix for a more detailed exposition.) Since $t_0^2=0$ there are associated cohomology groups $H_t^{p,q}$ and for the case in hand it is known that these vanish for $p>0$. Thus for $\ell >2$ the solution to \eq{3.5} will be the cohomologically trivial one

\be
I_{\ell-2,4}=t_0 J_{\ell-1,2}\ .
\la{3.6}
\ee

Now the explicit form of $I_{\ell-2,4}$ is

\be
I_{\ell-2,4}=t_0 F_{\ell -1,2} +\sum_{m+n=\ell} (F_{m+1}\wedge F_{n+1})_{\ell-2,4}\ ,
\la{3.9}
\ee

and so \eq{3.6} guarantees that the second term on the right-hand side is itself $t_0$ exact. This means that setting $J_{\ell-2,4}=0$ allows one to solve for $F_{\ell-1,2}$ in terms of the components of the lower-degree $F$s up to a $t_0$ exact term that can be absorbed by a redefinition of the potential $A_{\ell}$, where $F_{\ell+1}= d A_{\ell} + \ldots$.  This argument can be repeated for the higher-dimensional components of $I_{\ell + 2}$ and then for the higher degree forms sequentially. We therefore conclude that the entire system of Bianchi identities for the forms will be consistent provided that \eq{3.4} is satisfied and that the Bianchi identities for $F_2$ and $F_3$ have been solved. The $F_2$ solution was given previously. To see that there is a problem for $F_3$ in the Weyl representation we note that that the dimension-zero component of the Bianchi identity is

\be
t_0 F_{1,2}= F_{0,2}\wedge F_{0,2} \ .
\la{3.10}
\ee

When $F_2\wedge F_2$ is taken in the Weyl representation one can see that the right-hand side of this equation does not vanish and is certainly not $t_0$-exact. Thus this representation must be excluded.

The set of all forms together with their consistent Bianchi identities constitutes a Lie super co-algebra. We recall that this is a $\bbZ_2$-graded vector space $\cA$ together with a linear map $d:\cA \rightarrow \wedge^2 \cA$ which extends to a degree-one graded derivation of the graded exterior algebra $\wedge^* \cA$ and which squares to zero, where $\wedge$ denotes the graded antisymmetric tensor product. In our case there is also a $\bbZ$-grading,

\be
\cA=\oplus_{\ell\in \bbZ,\, \ell\geq 1} \cA_{\ell}=\cA^+ \oplus \cA^-
\la{3.11}
\ee

where $\cA_{\ell}$ is the space of $(\ell+1)$ field-strength forms, and where $\cA^+$ and $ \cA^-$ denote  the even and odd parts corresponding to $\ell$ even and $\ell$ odd respectively.

In subsection 3.3 we shall discuss the structure of the Lie superalgebra which is dual to this co-algebra, but before that we give a more detailed exposition of the field-strength forms up to degree five.


\subsection{Example of a consistent Bianchi identity}


The Bianchi identity for the five-form field strength transforming under the {\tiny$ \Yvcentermath1 \yng(2,1,1)$}-representation of $SO(8,n)$ is

\bea \label{3.3}
dF_5{}^{[MNO],P}&=&m(\bar{F}_4{}^{[MN}F_2{}^{O]P}+\bar{F}_4{}^{P[M}F_2{}^{NO]}+\frac{4}{6+n}\h^{P[M}\bar{F}_4{}^N{}_QF_2{}^{O]Q})\nm
&+&p(F_4{}^{P[M}F_2{}^{NO]}-\frac{2}{6+n}\h^{P[M}F_4{}^N{}_QF_2{}^{O]Q})\nm
&+&q(F_4{}^{MNO}{}_QF_2{}^{PQ}+F_4{}^{P[MN}{}_QF_2{}^{O]Q}-\frac{4}{6+n}\h^{P[M}F_4{}^{NO]}{}_{QR}F_2{}^{QR})\nm
&+&r(F_4{}^{MNO,}{}_QF_2{}^{PQ}+F_4{}^{P[MN,}{}_QF_2{}^{O]Q}-\frac{4}{6+n}\h^{P[M}F_4{}^{NO]}{}_{Q,R}F_2{}^{QR})\nm
&+&s(F_4{}^{[MN}{}_{Q,}{}^{O]}F_2{}^{PQ}+F_4{}^{P[M}{}_{Q,}{}^NF_2{}^{O]Q}+\frac{1}{6+n}\h^{P[M}(3F_4{}^{N}{}_{QR,}{}^{O]}+F_4{}^{NO]}{}_{Q,R})F_2{}^{QR})\nm
&+&t(F_4{}^{MNO}{}_{QR,}{}^PF_2{}^{QR}+F_4{}^{P[MN}{}_{QR,}{}^{O]}F_2{}^{QR})\nm
&+&u(F_3{}^{[MN}{}_{QR}F_3{}^{O]PQR} +\frac{2}{6+n}\h^{P[M}F_3{}^N{}_{QRS}F_3{}^{O]QRS})\nm
&+&v(F_3{}^{MNO}{}_VF_3{}^{PV}-F_3{}^{P[MN}{}_{V}F_3{}^{O]V})
\eea

where $M,N,O,P$ are $SO(8,n)$ vector indices and $m,p,..,v$ are real constants. If the Bianchi identity is consistent the constants can be chosen such that $ddF_5{}^{[MNO],P}=0$. If this can be done in $n$-ways the form is $n$-fold degenerate. If the coefficients cannot be chosen such that $ddF_5{}^{[MNO],P}=0$ the Bianchi identity is inconsistent and the form field transforming under this representation will not be a part of the form-field spectrum.

If we take the exterior derivative of \eqref{3.3} we obtain terms of the form $F_3{}^X\wedge F_2 \wedge F_2$ where $X$ indicates any one of the representations that the three forms transform under (these are listed in appendix \ref{forms}). The terms that are non-zero are those for which {\tiny$ \Yvcentermath1 \yng(2,1,1)$} is contained in the direct product $\Yvcentermath1 X \otimes $ {\tiny $ \Yvcentermath1 \yng(1,1) $} $\otimes $ {\tiny $ \Yvcentermath1 \yng(1,1)\  $}. The consistency of the Bianchi identity  gives five equations involving the constants $m,p...,v$; they are soluble provided that

\bea \label{d}
r&=&q-\frac{2}{3}u+\frac{6(2+n)}{5(4+n)}t\nm
s&=&-2q+2u-\frac{6n}{5(4+n)}t\nm
27m&=&\frac{1}{3}(4+n)u+\frac{3}{5}(4+n)t+2v\nm
3p&=&4v-4r+s\, .
\eea

Using the cohomological argument given in the previous section we know that there exists a solution to this Bianchi identity since it is consistent. There are eight unknown constants in the Bianchi identity and four constraints. We can therefore conclude that five-forms transforming under the {\tiny $ \Yvcentermath1 \yng(2,1,1)$}-representation of $SO(8,n)$ are allowed by supersymmetry and that there  is a fourfold degeneracy.

\subsection{Form Fields}

We give the Bianchi identities for all two-, three-,  and four-forms and for a few of the five-forms in the appendix \ref{forms}. The allowed form fields of degree $\geq 4$ and their degeneracies are derivable from group theory alone, but we have verified this explicitly for all four-forms and for five of the five-forms.  The possible form fields of degree $\leq 4$ where first presented in \cite{Bergshoeff:2008} where a Kac-Moody approach was used. We have re-derived these results and extended them to include the five-forms assuming only supersymmetry. Some of the five-forms are non-zero in supergravity and are also needed in the complete gauged theory. Our results are presented in table 1.

\begin{table}[ht] \label{yrepresentations}
\centering 
\begin{tabular}{|c | c || c |     c|} 
\hline
Form degree & Allowed forms & Form degree & Allowed Forms  \\ [0.5ex] 
\hline \hline
&&&\\
2 & {\tiny $ \Yvcentermath1 \yng(1,1)$} & 4 & {\tiny$ \Yvcentermath1  \yng(1,1)\ \ \yng(1,1,1,1)\ \ \yng(2)\ \  \yng(2,1,1)\ \ \yng(2,1,1,1,1)$} \\
&&&\\
\hline
&&&\\
3 & $\Yvcentermath1 1 $ {\tiny $\Yvcentermath1  \  \yng(1,1,1,1)\ \ \yng(2)$} & 5 & {\tiny $\Yvcentermath1  4 \cdot \yng(1,1)\ \ 2 \cdot \yng(1,1,1,1)\ \ 2 \cdot \yng(1,1,1,1,1,1)\ \ \yng(2)\ \  4 \cdot \yng(2,1,1)\ \ 2\cdot\yng(2,1,1,1,1)$}  \\
 && & {\tiny $\Yvcentermath1  \yng(2,1,1,1,1,1,1)\ \  2 \cdot \yng(2,2,1,1)\ \ \yng(2,2,2,1,1)\ \ \yng(3,1)\ \ \yng(3,1,1,1)\ \ \yng(3,1,1,1,1,1)$}  \\
&&&\\
\hline 
\end{tabular}
\caption{The SO(8,n) representations and their degeneracy for forms of degree $\leq$ 5.}
\label{table:nonlin} 
\end{table}

It is straightforward to construct the possible non-zero components of forms of any degree up to coefficients that can be determined by the Bianchi identites up to overall normalisation. These components can have dimension zero (two odd indices), one-half (one odd index) or one (no odd indices), so five-forms can only be non-zero at dimension zero, i.e. the $F_{3,2}$ components, while four-forms can  have non-zero dimension-zero and one-half components, $F_{2,2}$ and $F_{3,1}$ respectively. Clearly forms with degree higher than five must be identically zero in supergravity. The possible non-zero components have scalars, spinors, or dimension-one fields times appropriate invariant tensors. 

As an example, consider the four-form in the adjoint representation; its non-zero components are

\bea
            F_{ a b \a I\b J}{}^{RS}&=&ia(\c_{ab})_{\a\b}(\S^{ij})_{IJ}V_i{}^R V_j{}^S\nm
            F_{ a b c \a I}{}^{RS}&=&ib\vare_{abc}(\S^i\L^r)_{\a I} V_r{}^{[R} V_i{}^{S]},
\eea

where $a,b$ are  real constants. The only non-zero component of the five-form in the adjoint representation is

\bea
            F_{ abc \d I \e J}{}^{RS}&=&ic\vare_{abc}\vare_{\d\e}(\S^{ij}){}_{IJ}V_i{}^R V_j{}^S\ ,
\eea
 
for $c$ another real constant. In general, a five-form will be zero unless the representation of $SO(8,n)$ under which it transforms contains the adjoint representation of $SO(8)$.


\section{Borcherds Algebra}


The form fields were first given an algebraic interpretation in \cite{Cremmer:1998px} where a generator was associated to each potential such that the Maurer-Cartan equation for the sum of all field strengths generates the field equations. Two years later a correspondence between toroidal compactifications of M-theory and del Pezzo surfaces was found in \cite{Iqbal:2000}. Studying the cohomology of the del Pezzo surfaces the authors of \cite{HenryLabordere:2002dk} managed to extract the algebras found in \cite{Cremmer:1998px}. These algebras are Borcherds algebras and a truncated set of their positive roots correspond to the generators of the potentials. The set of roots also contained information about the deformation and top form potentials. 
 
The discrepancy bewteen the fact that there are infinitely many positive roots while there are only finitely many form fields in supergravity can be resolved in superspace, where forms can have any degree since the odd basis forms commute. In supergravity there are no non-zero field-strength forms of degree greater than $D+2$. However,  these forms could in principle become non-zero when we allow for corrections at order $\a'$. In \cite{Greitz:2011da} it was shown that the form fields in type $IIA$ and $IIB$ supergravity of degree larger than space-time are correctly encoded by Borcherds algebras.

The form fields of half-maximal supergravities have previously been given by the authors of \cite{Bergshoeff:2008}. These were found by constructing the branching of an extended version of the duality group, $G_D$, called $G_D{}^{+++}$, \footnote{$G_D{}^{+++}$ refers to the very extension of the duality group of the supergravity theory that has been dimensionally reduced to three dimensions.} with respect to $SL(D,\bbR)\xz G_D$. This decomposition includes modules corresponding to all the physical states, the deformation- and top-form potentials, as well as other states that currently have no interpretation. The modules corresponding to the form fields forms a truncated algebra called the $p$-form algebra. However, in terms of the $SL(D,\bbR)\xz G_D$ decomposition of $G_D{}^{+++}$ the degree of the form fields will not exceed the space-time limit.


\subsection{The Borcherds algebras for half-maximal supergravity}


We saw in section 3.1 that the forms in a supergravity theory, provided that their Bianchi identities are consistent and soluble, form a Lie super co-algebra. The consistency conditions for the Bianchi identities encode the Jacobi identity for the dual Lie superalgebra. Since the forms can have arbitrarily high degrees, this algebra is infinite-dimensional.

To find the Lie superalgebra dual to the co-algebra we shall make use of the techniques used in \cite{Damour:2002} to decompose $E_{10}$ in a level-by-level expansion. The form fields in supergravity should be in one-to-one correspondence with the positive roots of an infinite Lie superalgebra. We will make the assumption that a positive root of the Lie superalgebra can be written as

\be \label{root}
\a=\ell\a_0+\sum m^j\a_j\ ,
\ee

where $\ell, m^j$ are positive integers denoting the number of times the simple roots $\a_0$ and $\a_j$ appear in $\a$, and where $\ell$, which corresponds to the degree of the potential form in question, will be referred to as the level of the root. We label the generators associated to the simple roots $\pm \a_0$ and $\pm \a_i$ by $e_0\,(f_0)$ and $e_i\,(f_i)$. In addition, there will be generators of the Cartan subalgebra $(h_0,h_i)$. We will take the generators $e_i$ and $f_i$ to be the generators of the duality group for reasons that will soon become clear, so that the index $i$ runs from 1 to the rank of the duality group $SO(8,n)$. The level of a generator is given by the number of times $e_0$ appears in its expression as a (multiple) commutator of level-one generators. The adjoint action of $e_i$ on a generator does not alter its level so that all generators at a given level transform under a direct sum of representations of the duality group.

The representations appearing at level $\ell +1$ are contained in the product $\cR_{\ell}\otimes \cR_1$, where $\cR_{\ell}$ denotes the representations appearing at level $\ell$ and $\cR_1$ is the adjoint representation. Of the representations that do appear there will be one generator associated to its highest weight. This is most easily seen if one considers the generator $f^\L$ corresponding to the negative root $-\a$. If ad$e_i(f^\L)=0$, then $f^\L$ acts as a highest weight state for one of the representations appearing at level $\ell$. The weight of the state $f^\L$ is $h_i(f^\L)=p_i$, where $p_i$ is the Dynkin label for the representation. All generators that do not correspond to highest weight states are derivable from these by acting on $f^\L$ by ad$f_i$.

Determining the Cartan matrix $A$ for the Lie superalgebra is rather trivial given the above assumption. $A$ will be completely specified by analysing the Bianchi identities for the two- and three-form Bianchi identities. The first step to note is that all generators appearing at each level must, in order for the duality to work, transform under representations of $SO(8,n)$. If $A$ has the form

\[A= \left( \begin{array}{cc}
A_{00} & A_{0i}   \\
A_{i0}  & A_{ij}
\end{array} \right),\]

as suggested by \eqref{root} then $A_{ij}$ is the Cartan matrix for $SO(8,n)$. To determine $A_{0i}$ we note that $f_0$ acts as a highest weight state for $SO(8,n)$ at level one. The weight of this state is $h_i(f_0)=-A_{i0}=p_i$. The two-form field strength is in the adjoint representation with Dynkin labels $(010...0)$, hence we demand the generators at level one to transform under the same representation. We can therefore conclude that $A_{0i}=(0,-1,0,...,0)$ and without loss of generality we can take $A_{i0}=A_{0i}$.

To determine $A_{00}$ we will match the representations at level two in the roots to those appearing in the three-forms. The generators at level two are formed by commuting the generators at level one, and the representations that can appear at level two are therefore contained in the symmetric product

\Yvcentermath1
\be
\left({\tiny \yng(1,1)\otimes \yng(1,1)}\right)_S = {\tiny \yng(2,2) + \yng(1,1,1,1) +\yng(2) +1}.
\ee

$A_{00}$ could take the values $\leq -1$, $0$ or $2$; the roots corresponding to these values were given the following Dynkin diagrams in \cite{HenryLabordere:2002dk}  

\vspace{1cm}
\begin{center}
\begin{tabular}{|l|l|}
\hline
\begin{picture}(14,14)
\thicklines
\put(7,4){\circle{14}}
\end{picture}
&
Bosonic real root of length 2
\\
\hline
\begin{picture}(14,14)
\thicklines
\put(7,4){\circle{14}}
\put(2,-1){\line(1,1){10}}\put(2,9){\line(1,-1){10}}
\end{picture}
&
Bosonic imaginary root of length $\leq$ 0
\\
\hline
\begin{picture}(14,14)
\thicklines
\put(7,4){\circle*{14}}
\end{picture}
&
Fermionic ``imaginary'' root of length 1, $A_{00}=0$ 
\\
\hline
\begin{picture}(14,14)
\thicklines
\put(7,4){\circle{14}}
\put(7,4){\circle*{10}}
\end{picture}
&
Fermionic imaginary root of length $\leq$ -1 \\
\hline
\end{tabular}
\end{center}
\vspace{1cm}

We will discuss the different nodes in turn. If $A_{00}\leq -1$ then $[f_0,f_0]$ is a generator at level two. Moreover, it would be a highest weight state of $SO(8,n)$ since ad$e_i([f_0,f_0])=0$. This generator would therefore give rise to the Weyl-tensor representation appearing at level two with weight $h_i([f_0,f_0])=(020...0)$. Going back to the form fields we see that this representation is not allowed by supersymmetry so we cannot choose $A_{00}=-1$. If $A_{00}=2$ only the adjoint representation appears at level 3. Hence we are left with $A_{00}=0$. There are two type of nodes with length $0$, bosonic or fermionic however $e_0$ need to be fermionic to reflect that the two forms commute, leaving us with the following Borcherds algebras

\pagebreak
\begin{picture}(200,60)(0,0)
\thicklines
\multiput(60,-20)(42,0){4}{\circle{14}}
\put(102,22){\circle*{14}}
\put(67,-20){\line(1,0){28,0}}
\put(102,-13){\line(0,1){29}}
\put(109,-20){\dashbox{2}(28,0)}
\put(149,-15){\line(1,0){32,0}}
\put(149,-25){\line(1,0){32,0}}
\put(160,-30){\line(1,1){10,10}}
\put(170,-20){\line(-1,1){10,10}}
\multiput(272,-20)(42,0){3}{\circle{14}}
\put(314,22){\circle*{14}}
\put(387,11){\circle{14}}
\put(388,-50){\circle{14}}
\put(279,-20){\line(1,0){28,0}}
\put(314,-13){\line(0,1){29}}
\put(321,-20){\dashbox{2}(28,0)}
\put(362,-15){\line(1,1){20,20}}
\put(382,-45){\line(-1,1){20,20}}
\put(75,-65){Figure 1.}
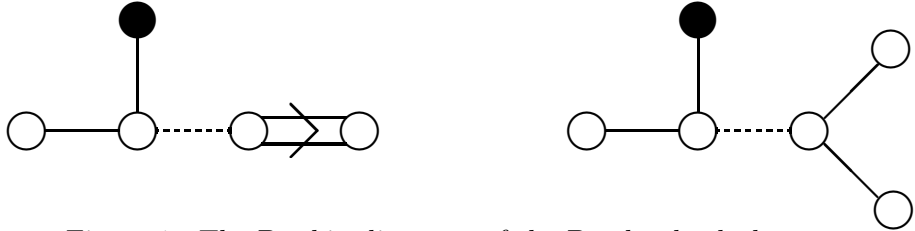
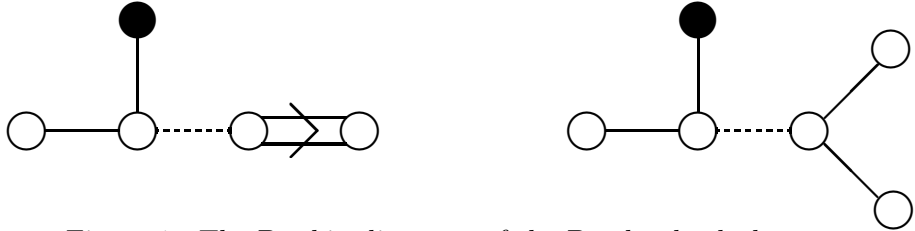
\put(125,-65){The Dynkin diagrams of the Borcherds algebras}
\put(125,-77){encoding the form field strengths}
\end{picture}
\\
\\
\\
\\
\\
\\

The Dynkin diagrams in figure 1 correspond to the Borcherds algebras that encode the form field spectrum. When the duality group is $SO(8,2n-1)$ the diagram to the left is relevant, while if the duality group is $SO(8,2n)$ the Dynkin diagram to the right should be used. We have verified that the above Borcherds algebras do indeed reproduce the representations in table 2. We have done this by using a generalisation of the result from \cite{Henneaux:2010ys}, \cite{Palmkvist:2011vz} that the $p$-form spectrum of $E^{+++}$ is a truncated Borcherds algebra. The generalization given in \cite{Palmkvist:2012} states that the level decomposition with respect to a fermionic simple root of length zero in a Borcherds algebra can be obtained by replacing the corresponding black node with an infinite chain of white nodes, corresponding to bosonic simple roots of length 2. The upshot of this is that we can use a computer program \cite{Nutma} to calculate the representations up to any level by adding appropriately many white nodes to a $B_n$ or $D_n$ diagram to find the representation content at each level. The method of decoding the information of the Cartan matrices defined by the Dynkin diagrams in figure 1 is thus equivalent to the way the authors of \cite{Bergshoeff:2008} found the allowed form fields, the difference being that one adds more white nodes if one is interested in form fields of higher degree. The modules that do not go with totally antisymmetric tensors are not defined by the Bianchi identities. From this point of view they are objects appearing when one extracts the representation content at each level from the Borcherds algebra using a Kac-Moody algebra.  

The above analysis is not limited to three dimensions nor to the particular duality groups $SO(8,n)$. It will also be the case that Borcherds algebras are defined by the Bianchi identities in other supergravity theories. An example was given in \cite{Greitz:2011da} where type $IIA$ and $IIB$ supergravity was analysed in a similar manner.


\section{Corrections at order $\a'$}


In the presence of corrections of order $\a'$ some higher-degree forms can in principle have non-zero components. Forms with bi-degrees $(p,6)$ can have contributions of the form $\a'$ times scalars, while $(p,5)$-forms can have contributions linear in $\L$ multiplied by $\a'$. Here we shall focus on the latter as they are slightly easier to discuss. In principle this could be affected by neglecting the former, but for the non-trivial example to be discussed below it will turn out that there can be no such contribution. To simplify things we shall also consider only the case $n=1$, i.e. the duality group is $SO(8,1)$.

The Bianchi identities we need to consider have the form

\be
d F_n= F_2\wedge F_{n-1} + \ldots\ ,
\la{6.11}
\ee

where $n=p+5$ for $p=0,1,2,3$. In particular we shall focus on the case $p=1$ and make the assumption that $F_{0,6}=0$. The lowest non-trivial component of \eq{6.11} that we are interested in has the form

\be
t_0 \stackrel{(1)}{F_{1,5}}= \stackrel {(0)}{F_{0,2}} \stackrel {(1)}{ F_{0,5}}\ ,
\la{6.12}
\ee

since no other terms can contribute at order $\a'$. Here, the superscripts indicate the order of $\a'$ in the given terms. As a first example, let us consider the case when the six-form is in the adjoint representation. The Bianchi identity is

\be
dF_6^{RS}= F_2^{T[R} F_5^{S]}{}_T + \ldots\ .
\la{6.13}
\ee

We can now use the scalar matrix to rewrite this equation in an $SO(8) \xz SO(1)$ basis. For the term we are interested in this will be valid provided that there is no scalar contribution in $F_{0,6}$. We therefore find a term

\be
t_0 F_{1,5}^i= F_{0,2}^{ij} F_{0,5\,j}\ ,
\la{6.14}
\ee

where here, and below, we omit the order superscripts as it should be clear from the context which ones are meant. This is the only term that can appear on the right because $F_{0,5}$ has an odd number of unprimed $Spin(8)$ indices, so that we need an odd number of external vector indices in order to be able to find a linear $\L$ term. The $F_2$ term is 

\be
F_{\a I\b J}^{ij}= i \ve_{\a\b} (\S^{ij})_{IJ}\ .
\la{6.15}
\ee

The $F_{0,5}$ term must contain the spinor $\L^{I'}_\a$, and since the five odd indices are totally symmetric, it follows that the $Spin(8)$ indices must be in the Young tableau arrangement $\Yvcentermath1\tiny\yng(3,2)$. This decomposes into the following representations

\be
\Yvcentermath1{\tiny{\yng(3,2)}}=(0210)+(0030)+(0110) + (0010)\ .
\la{6.16}
\ee

We need to multiply these by the additional vector index, or $(1000)$, and then look for possible $(0001)$s which could correspond to the spinor field $\L$. There is just one possibility and that comes from the $(0010)$ representation in \eq{6.16}. Before we examine the right-hand side it is necessary to check whether this possibility is trivial in the sense that it could be removed by a field redefinition of the potential $A_{1,3}^i$. Consider the sequence

\be
\O_{2,1}^i \stackrel{t_0}{\longrightarrow}\O_{1,3}^i \stackrel{t_0}{\longrightarrow}\O_{0,5}^i \ ,
\la{6.17}
\ee

where $\O_{p,q}^i $ denotes the space of $(p,q)$-forms with an additional vector index $i$. If the element we are interested in is the image of $t_0$ acting on $\O_{1,3}^i $ then it can be removed by a field redefinition. Now there is just one possible $\L$ term in $\O_{2,1}^i$, namely

\be
(\c_{ab}\S^i \L))_{\a I}\nn\ ,
\ee

while there are two possible $\L$ terms in $\O_{1,3}^i $, as one can see by a little group-theoretical analysis. (In this case there are two possible arrangements of the Lorentz spinor indices due to the additional Lorentz vector index, so the $Spin(8)$ indices can be in the tableaux $\Yvcentermath1\tiny\yng(2,1)$ or $\tiny\yng(3)$.) One of these must therefore be $t_0$ exact, so the second one must map to $\O_{0,5}^i $. If this were not the case, this element would have to be in the cohomology group $H_t^{1,3}$, but this is zero. The conclusion of this analysis is that there are no non-trivial $\L$ terms in the six-forms in the adjoint representation. 

It turns out that a similar situation obtains for the six-forms in the smallest representations of $SO(8,1)$, i.e. $((0000),(0100), (1000),(2000)$ and $(0010)$, so that the first representation that can provide a non-trivial solution is in the four-form representation of $SO(8,1)$, i.e. $(0002)$. 

The Bianchi identity is

\be
d F_6^{MNPQ}= F_2^{RS} F_5^{MNPQ,}{}_{RS} + \ldots\ ,
\la{6.18}
\ee

where $F_5$ on the right is in the $\Yvcentermath2\tiny\yng(2,2,1,1)$ representation.
Projecting onto $SO(8)$ indices, we find that there are two possible $SO(8)$ representations for $F_5$ on the right-hand side that can contain $\L$ given by the tableaux $\Yvcentermath2\tiny\yng(2,2,1)$  and $\Yvcentermath2\tiny\yng(2,1,1,1)$, or $(0111)$ and either $(1020)$ or $(1002)$ in terms of $SO(8)$ Dynkin labels. It turns out that both the latter cannot contain any non-trivial $\L$ terms and so can be discarded. The relevant term in the Bianchi identity  is therefore

\be
t_0 F^{ijk}_{1,5}=F_{0,2}^{lm} F_{0,5}^{ijk,}{}_{lm}
\la{6.19}
\ee

The analysis goes in the same way as the previous example. The $Spin(8)$ indices on $F_{0,5}$ are again in the $\Yvcentermath1\tiny\yng(3,2)$ tableau, while the additional $SO(8)$ indices are in the representation $(0111)$. We find there are two possible $\L$ terms but that one of them is $t_0$ exact and so can be removed by a field redefinition. So the question is whether this term, when multiplied by $F_{0,2}$, becomes $t_0$ exact.  To answer this consider the sequence

\be
\O_{3,1}^{ijk}\stackrel{t_0}{\longrightarrow}\O_{2,3}^{ijk} \stackrel{t_0}{\longrightarrow}\O_{1,5}^{ijk} \stackrel{t_0}{\longrightarrow}\O_{0,7}^{ijk}\ .
\la{6.20}
\ee

It is straightforward to find the number of possible $\L$ terms that can occur in each space. We find $1,3,4$ and $2$ such terms in each space starting from the left. Since there is no $t_0$ cohomology except perhaps for $H_t^{0,7}$, we can immediately see that there can be two non-trivial $\L$ terms in $F_{1,5}^{ijk}$ and therefore both of the $\L$ terms in the $(0,7)$ form,  $J^{ijk}_{0,7}$ say,  are in fact in the image of $t_0$. In other words, $H_t^{0,7,ijk}$ restricted to the representation $(0001)$ vanishes. As we have seen there are two possible $\L$ terms in $F_{0,5}^{ijk,}{}_{lm}$ and these give rise to the two $\L$ terms in $J_{0,7}^{ijk}$. In fact, the $Spin(8)$ spinor indices for $J_{0,7}^{ijk}(\L)$ must be in the tableau $\Yvcentermath2\tiny\yng(4,3)$ which can be rewritten as

\be
{\Yvcentermath2\tiny\yng(4,3)=\tiny\yng(3,2) }+ (0310) + (0130)\ .
\la{6.21}
\ee

When one tensors this with the representation $(0011)$ one finds that the last two representations cannot give rise to a $\L$, whereas the first gives rise to two possibilities of this type. Clearly these correspond to the two we have identified earlier in $F_{0,5}^{ijk,}{}_{lm}$. So there is a single non-trivial solution to this Bianchi identity.

It is not difficult to see that this conclusion cannot be affected by a possible scalar term in $F_{0,6}$ on the left. This could give a term of the form $d_1 F_{0,6}^{MNPQ}$, projected onto an $SO(8)$ basis. In order to have a Lorentz scalar in $F_{0,6}$ the $Spin(8)$ indices would have to be in the $\Yvcentermath2\tiny\yng(3,3)$ tableau, but the four $SO(8,1)$ indices, which are totally antisymmetric, give rise to at least three antisymmetrised indices when broken down to $SO(8)$ representations and therefore a scalar term cannot be accommodated because they would need to be contracted with three antisymmetrised indices coming from the odd form indices.


\section{Conclusions}


In this paper we have used supergeometrical methods to analyse various $D=3, N=8$ theories starting from the superspace constraints that correspond to off-shell conformal supergavity. The Poincar\'e theories were constructed using standard coset methods for the sigma models $(SO(8) \xz SO(n)\bsh SO(8,n)$. The constraints were modified to incorporate general gaugings and it was noted that this can only be done if one uses the non-minimal conformal constraints.

In the rest of the paper we focused on the algebraic structure of the ungauged Poincar\'e theories. The set of all possible forms with consistent, soluble Bianchi identities were shown to define a Lie super co-algebra and we were able to identify the dual Lie algebra with the positive sector of Borcherds algebras formed by adding a single odd root to the root system of the duality algebra $\gs\go(8,n)$. The analysis is made extremely simple by the use of superspace cohomology which one can use to show that the consistent Bianchi identities for all forms with degree greater than three are automatically satisfied. We were also able to relate the forms constructed in this way with those of infinitely extended Lie algebras.

In a superspace setting the field strength forms can be non-zero in supergravity up to degree five, and we indicated how the Bianchi identities can be solved for such forms. In the presence of $\a'$ corrections one might expect that higher-degree forms might be turned on and we gave a simple example of this for certain six-forms. It should be emphasised, however, that this is only a very partial analysis. In principle we should go back to the beginning and solve all of the Bianchi identities sequentially and we are not guaranteed a priori that this can be done. Were it to be the case that this system of forms is not consistent in the presence of higher-order corrections then it would mean that the Borcherds algebra picture would be restricted to the supergravity limit.

A final comment on corrections is that, in $N=1,D=10$ supergravity, it is well-known that there are $\a'$ corrections to the Bianchi identities for the three-forms. These are required for anomaly cancellations, but such considerations should not be important in three dimensions. However, they are possible on dimensional grounds and should they occur they would also interfere with the algebraic structure. It should not be too difficult to compute all the corrections at order $\a'$. Such a computation, although lengthy, would be a useful thing to carry out in order to check whether the algebraic picture given here survives at this order.

\pagebreak

{\bf\Large{Acknowledgements}}

We would like to thank Nordita and the organisers of the programme ``Geometry of strings and fields" where part of this work was carried out. We also would like to thank the organisers of the 28th Nordic network meeting on strings, fields and branes where we met Jakob Palmkvist whom we thank for interesting and useful discussions. JG thanks Tekn. Dr Marcus Wallenbergs Stiftelse and the STFC for financial support. 

\vskip .5cm

 \appendix
 
 {\bf \Large{Appendices}}
 
 
\section{Spacetime coventions}


The metric is $\h_{ab}={\rm diag}(-1,1,1)$. The epsilon tensor is defined so that $\ve_{012}=+1$. The dual of a one-form $v_a$ is $v_{ab}:=\ve_{abc} v^c$ so that $v_a=-\frac{1}{2}\ve_{abc} v^{bc}$.

The gamma-matrices with indices in standard position are $(\c^a)_\a{}^\b$. They obey the algebra $\c_a \c_b =\h_{ab} + \c_{ab}$, where $\c_{ab}=\ve_{abc}\c^c$. We also have $\c_{abc}=\ve_{abc}$ for the totally antisymmetrised product of three gamma-matrices. Spinor indices are lowered or raised with the spin ``metrics'' $\ve_{\a\b}$ and $\ve^{\a\b}$ which we take to have the same numerical entries, i.e. $\ve_{12}=\ve^{12}=+1$. The summation convention is NE-SW, i.e. $v^\a=\ve^{\a\b} v_\b$ and $v_\a=v^\b\ve_{\b\a}$. The matrices $\c_a$ (and $\c_{ab}$) with both spinor indices down (or up) are symmetric.

A vector can be written as a symmetric bi-spinor via

\be
 v_{\a\b}=-\half (\c_a)_{\a\b} v_a \Leftrightarrow v_a= (\c_a)^{\a\b} v_{\a\b} \   .
 \la{a1}
\ee

For any two spinors $\psi,\chi$ and any gamma-matrix $\C$ we define the tensorial bilinear to be

\be
 \psi\C\chi:= \psi^\a \C_\a{}^\b \chi_\b\  .
 \la{a2}
\ee

 
\section{Conventions for $SO(8)$ and $SO(8,n)$}.


$SO(8)$ vector indices are $i,j,\ldots =1\ldots 8$, unprimed Weyl spinor indices are $I,J,\ldots= 1\ldots 8$ and primed Weyl spinor indices are $I',J',\ldots =1\ldots 8$.  These correspond to the representations (1000), (0010) and (0001), respectively. The metrics for each three spaces are flat euclidean, so it is not important to distinguish between upper and lower indices.

The basic sigma-matrices are $(\S_i)_{IJ'}$ and $(\tilde\S_i)_{J'I}$. We shall take $\tilde\S_i=(\S_i)^T$ and not bother to write out the tildes since it will be clear from the context which is meant. Sigma-matrices with two or more indices are antisymmetrised products of the basic ones as usual.

Sigma-matrices with an even number of vector indices are bi-spinors of a fixed chirality. $\S_2$ give a basis of antisymmetric $8\xz 8$ matrices while $(1,\S_4)$ give  basis of symmetric matrices. We shall take $(\S_{i_1\dots i_4})_{IJ}$ to be self-dual while $\S_4$ with primed indices is anti-self-dual.

For an arbitrary matrix $M_{IJ}$ we have

\be
 M_{IJ} =\frac{1}{8} \sum_{n=0}^{n=2} (\S^{i_1\dots i_{2n}})_{IJ} M_{i_1\dots i_{2n}}\ ,
 \la{a3}
\ee

where 

\be
 M_{i_1\dots i_{2n}}:= \frac{1}{(2n)!}(\S_{i_1\dots i_{2n}})^{IJ} M_{IJ}\ ,
 \la{a4}
\ee

except for $n=4$ when there is an extra factor of $\half$ on the right-hand side. The matrix $\S_0$ is  $\d_{IJ}$. The formula for primed indices is identical.

The bilinears that can be formed from the spinor field $\L^r_{\a I'}$  in the text are the Lorentz scalars

\bea
B&=& \L\L:= \L^{\a I' r} \L_{\a I' r} \nn\w1
B_{i_1\ldots i_4}&=& \L\S_{i_1\dots i_4}\L:=\L^{\a I' r} (\S_{i_1\dots i_4})_{I'J'} \L_{\a J' r}\ ,
\la{a5}
\eea

and the spacetime vectors

\be
A_{a ij}=\L \S_{i_1 i_2}\c_a \L:=\L^{\a I'r} (\S_{ij})_{I'J'} (\c_a)_\a{}^\b \L_{\b J'r}  
\la{a6}
\ee

Vector indices for $SO(8,n)$ are denoted by $R,S$, etc, while those for $SO(n)$ are $r,s$, etc. Indices for the adjoint representation are denoted by $X,Y$, etc, so that $X=[RS]$. The metric is $\h_{RS}=(\d_{ij},-\d_{rs})$, and indices are raised and lowered using this metric, including $\h_{rs}$ for $SO(n)$ indices.


\section{Borcherds algebras}


The definition of a Borcherds (or generalised Kac-Moody) (super)-algebra starts with a generalised symmetric Cartan matrix, $(a_{ij}),\ i.j=1\dots N$, where some subset of the indices can be odd, which is non-degenerate and for which the following rules hold. The diagonal elements $a_{ii}$ (no sum) can be positive, negative or zero, while the off-diagonal elements, $a_{ij},\ i\neq j$, are less or equal to zero. In the case that $a_{ii}>0$, then $\frac{2a_{ij}}{a_{ii}}\in\bbZ, \forall j$, while if $i$ is also odd $\frac{a_{ij}}{a_{ii}}\in\bbZ, \forall j$.

The Borcherds algebra $\cA$ associated with $(a_{ij})$ is then determined by $3N$ generators $\{h_i,e_i,f_i\}$, $i=1\ldots N$, satisfying the following conditions:

\bea
[h_i,h_j]&=&0   \la{A.1}\w1
[h_i,e_j]&=&a_{ij} e_j,\qquad [h_i,f_j]=-a_{ij} e_j,\qquad [e_i,f_j]=\d_{ij} h_i \la{A.2}\w1
(\ad\, e_i)^{1-\frac{2a_{ij}}{a_{ii}}}e_j&=&0,\qquad {\rm for}\  \ a_{ii}>0\ {\rm and}\  i\neq j\la{A.3}\w1
[e_i,e_j]&=&0\qquad {\rm when} \ \ a_{ij}=0\la{A.4}\ ,
\eea

with the last two conditions remaining valid if $e_i,\,e_j$ are replaced by $f_i,\,f_j$. The generators $h_i$ are even, and the generator $f_i$ is even or odd if $e_i$ is. If $a_{ii}>0$ the integer $\frac{2a_{ij}}{a_{ii}}$ is negative, and if $i$ is odd, it is also even. 

In a Borcherds algebra there is still a triangular decomposition of the form $\cA=\cN^-\oplus\cH\oplus \cN^+$, and it is still possible to define roots as in the Kac-Moody case. Furthermore, if $a_{ii}>0$, the algebra generated by $\{f_i,h_i,e_i\}$ for $i$ even, or by these together with $[f_i,f_i]$ and $[e_i,e_i]$ when $i$ is odd, are isomorphic to $\gs\gl(2)$ or $\go\gs\gp(1|2)$, respectively, and the algebra can be decomposed into finite dimensional representations of these (super)algebras. When $a_{ii}<0$, one has the same algebras but the Borcherds algebra contains infinite-dimensional representations of them. In the case that $a_{ii}=0$, the sub-algebra generated by $\{f_i,h_i,e_i\}$ is isomorphic to the Heisenberg (super)algebra.

In the case of the Borcherds algebras encountered in half-maximal $D=3$ supergravities the standard forms for the Cartan matrices associated with the duality sub-algebras are not symmetric. However, they can be made so by multiplying them by appropriate diagnoal matrices in such a way as to ensure that the above conditions are valid.


\section{Superspace cohomology}\label{cohomology}


Since the tangent bundle splits into even and odd parts it is possible to split the space of $n$-forms into spaces of $(p,q)$-forms, $p+q=n$, where a $(p,q)$ form has $p$ even and $q$ odd indices:

\be
\O^{p,q}\ni \o_{p,q}=\frac{1}{p! q!} E^{\b_q}\ldots E^{\b_1} E^{a_p}\ldots E^{a_1} \o_{a_1\ldots a_p\b_1\ldots \b_q}\ ,
\la{c1}
\ee

where, in this appendix, spinor indices run from 1 to 32. The exterior derivative splits into four terms with different bidegrees:

\be
d=d_0+ d_1 + t_0 + t_1\ ,
\la{c2}
\ee

where the bidegrees are $(1,0),(0,1), (-1,2)$ and $(2,-1)$ respectively. The first two, $d_0$ and $d_1$, are essentially even and odd differential operators, while the other two are algebraic operators formed with the dimension-zero and dimension three-halves torsion respectively. In particular,

\be
(t_0 \o_{p,q})_{a_2\ldots a_p \b_1\ldots \b_q}\propto T_{(\b_1\b_2}{}^{a_1}\o_{a_1|a_2\ldots a_p|\b_3\ldots \b_{q+2})}\ .
\la{c3}
\ee

The equation $d^2=0$ splits into various parts according to their bidegrees amongst which one has

\bea
(t_0)^2&=& 0\la{c4}\w1
t_0 d_1 + d_1 t_0&=&0\la{c5}\w1
d_1^2 +t_0 d_0+ d_0 t_0&=&0\ .
\la{c6}
\eea

The first of these enables us the define the cohomology groups $H_t^{p,q}$, the space of $t_0$-closed $(p,q)$-forms modulo the exact ones \cite{Bonora:1986ix}. The other two then allow one to define the spinorial cohomology groups $H_s^{p,q}$, but we shall not need these in this paper. In ten and eleven dimensions these cohomology groups are related to spaces of pure spinors and pure spinor cohomology respectively  \cite{Howe:1991mf,Howe:1991bx,Berkovits:2002zk}.

In $D=3, N=16$ supergravity the dimension-zero torsion is given in equation \eq{2.1}. The associated $t_0$ turns out to have trivial cohomology for $p\geq 1$, a result that greatly simplifies the problem of finding solutions to the differential form Bianchi identities. It can be derived by dimensional reduction from $D=10$ \cite{Berkovits:2008qw} cohomology. It has also been discussed using different techniques in \cite{Brandt:2010fa,Movshev:2011pr}.

\section{Bianchi identities} \label{forms}

\subsection{Two Forms}
\be
dF_2{}^{[MN]}=0
\ee

\subsection{Three Forms}

\bea
dF_3&=&F_{2OP}F_2{}^{OP} \nm
dF_3{}^{(MN)}&=&F_2{}^M{}_QF_2{}^{NQ} -\frac{1}{8+n}F_{2OP}F_2{}^{OP} \nm
dF_3{}^{[MNOP]}&=&F_2{}^{[MN}F_2{}^{OP]} 
\eea

\subsection{Four Forms}
\bea
d\bar{F}_4{}^{[MN]}&=&F_3F_2{}^{MN}+\frac{3}{7}(6F_3{}^{[M}{}_QF_2{}^{N]Q}-9BF_3{}^{MN}{}_{OP}F_2{}^{OP})\nm
dF_4{}^{(MN)}&=&F_3{}^{(M}{}_QF_2{}^{N)Q}\nm
dF_4{}^{[MNOP]}&=&F_3{}^{[MNO}{}_QF_2{}^{P]Q}\nm
dF_4{}^{[MNO],P}&=&(F_3{}^{P[M}F_2{}^{NO]}
-\frac{2}{6+n}\h^{P[M}F_3{}^{N}{}_QF_2{}^{O]Q})\nm
&+&\frac{3}{4}(F_3{}^{MNO}{}_QF_2{}^{PQ}+F_3{}^{P[MN}{}_QF_2{}^{O]Q}
-\frac{4}{6+n}\h^{P[M}F_3{}^{NO]}{}_{QR}F_2{}^{QR})\nm
dF_4{}^{[MNOPQ],R}&=&F_3{}^{[MNOP}F_2{}^{Q]R}+F_3{}^{R[MNO}F_2{}^{PQ]}
+\frac{6}{4+n}\h^{R[M}F_3{}^{NOP}{}_VF_2{}^{Q]V}\nm
\eea

\subsection{Five Forms}

\bea
d\bar{F}_5{}^{[MN]}&=&a\bar{F}_4{}^{[M}{}_QF_2{}^{N]Q}+bF_4{}^{[M}{}_QF_2{}^{N]Q}\nm
&+&cF_4{}^{MN}{}_{PQ}F_2{}^{PQ}+hF_4{}^{MN}{}_{P,Q}F_2{}^{PQ}\nm
&+&fF_3{}^M{}_QF_3{}^{NQ}+gF_3{}^M{}_{PQR}F_3{}^{NPQR}
\eea 
Constraints
\bea 
-\frac{54}{6+n}a-3b-\frac{4(5+n)}{6+n}h+6f&=&0\nm
-\frac{162}{6+n}a+3c+\frac{3(10+n)}{6+n)}h+6g&=&0
\eea

\bea
dF_5{}^{[PQRSTU]}&=&hF_4{}^{[PQRS}F_2{}^{TU]}
+iF_3{}^{[PQR}{}_VF_3{}^{STU]V}\nm
&+&j(F_4{}^{[PQRST,}{}_VF_2{}^{UV]}+F_{4V}{}^{[PQRS,T}F_2{}^{U]V})
\eea 
Constraints 
\be h+2i-\frac{10+n}{5(5+n)}j=0 \ee

\bea \label{c} 
dF_5{}^{[MNO],P}&=&o(\bar{F}_4{}^{[MN}F_2{}^{O]P}+\bar{F}_4{}^{P[M}F_2{}^{NO]}+\frac{4}{6+n}\h^{P[M}\bar{F}_4{}^N{}_QF_2{}^{O]Q})\nm
&+&p(F_4{}^{P[M}F_2{}^{NO]}-\frac{2}{6+n}\h^{P[M}F_4{}^N{}_QF_2{}^{O]Q})\nm 
&+&q(F_4{}^{MNO}{}_QF_2{}^{PQ}+F_4{}^{P[MN}{}_QF_2{}^{O]Q}-\frac{4}{6+n}\h^{P[M}F_4{}^{NO]}{}_{QR}F_2{}^{QR})\nm 
&+&r(F_4{}^{MNO,}{}_QF_2{}^{PQ}+F_4{}^{P[MN,}{}_QF_2{}^{O]Q}-\frac{4}{6+n}\h^{P[M}F_4{}^{NO]}{}_{Q,R}F_2{}^{QR})\nm
&+&s(F_4{}^{[MN}{}_{Q,}{}^{O]}F_2{}^{PQ}+F_4{}^{P[M}{}_{Q,}{}^NF_2{}^{O]Q}+\frac{1}{6+n}\h^{P[M}(3F_4{}^{N}{}_{QR,}{}^{O]}+F_4{}^{NO]}{}_{Q,R})F_2{}^{QR})\nm
&+&t(F_4{}^{MNO}{}_{QR,}{}^PF_2{}^{QR}+F_4{}^{P[MN}{}_{QR,}{}^{O]}F_2{}^{QR})\nm
&+&u(F_3{}^{[MN}{}_{QR}F_3{}^{O]PQR} +\frac{2}{6+n}\h^{P[M}F_3{}^N{}_{QRS}F_3{}^{O]QRS})\nm 
&+&v(F_3{}^{MNO}{}_VF_3{}^{PV}-F_3{}^{P[MN}{}_{V}F_3{}^{O]V})
\eea
Constraints
\bea \label{d}
r&=&q-\frac{2}{3}u+\frac{6(2+n)}{5(4+n)}t\nm
s&=&-2q+2u-\frac{6n}{5(4+n)}t\nm
27o&=&\frac{1}{3}(4+n)u+\frac{3}{5}(4+n)t+2v\nm
3p&=&4v-4r+s.
\eea

\bea \label{f} 
dF_5{}^{(MNO),P}&=&p(F_4{}^{(MN}F_2{}^{O)P}\nm
&-&\frac{1}{8+n}(2\h^{P(M}F_4{}^N{}_QF_2{}^{O)Q}-\h^{(MN}(F_4{}^{O)}{}_QF_2{}^{PQ}+F_4{}^{|P|}{}_QF_2{}^{O)Q})))\nm 
&+&q(F_4{}^{P}{}_Q{}^{(M,N}{}_QF_2{}^{O)Q}\nm
&-&\frac{1}{2(8+n)}(2\h^{P(M}F_{4QR}{}^{N,O)}F_2{}^{QR}-\h^{(MN}(F_{4QR}{}^{O),P}F_2{}^{QR}+F_{4QR}{}^{|P|,O)}F_2{}^{QR})))\nm 
&+&r(F_3{}^{(MN}F_3{}^{O)P}-\frac{1}{2(6+n)}F_3{}^{P[MN}{}_{V}F_3{}^{O]V})
\eea

Constraints
\bea \label{g}
p&=&r=-\frac{q}{3}
\eea

\bea
dF_5{}^{[MNOP],}{}_{[QR]}&=&\Delta(F_4^{MNOP}F_{2QR}+F_{4QR}{}^{[MN}F_2{}^{OP]}
+2F_4{}^{[MNO}{}_{[Q}F_2{}^{P]}{}_{R]}\nm
&-&\frac{10}{4+n}\h^{[M}_{[Q}(F_4{}^{NOP]}{}_TF_{2R]}{}^T
+F_{4R]}{}^{NO}{}_TF_2{}^{P]T})\nm
&+&\frac{20}{(4+n)(5+n)}\h^{[M}_{\phantom{[}Q}\h^N_R
F_4{}^{OP]}{}_{TS}F_2{}^{TS})\nm
&+&\Gamma((F_4{}^{[MNO,}{}_{[Q}
+F_{4[Q}{}^{MN,O})F_2{}^{P]}{}_{R]}\nm
&+&(F_4{}^{[MN}{}_{[Q,R]}
-F_{4QR}{}^{[M,N})F_2{}^{OP]}\nm
&-&\frac{1}{(4+n)}\h_{[Q}^{[M}((F_4{}^{NOP]}{}_T
+F_4{}^{[NO}{}_{T,}{}^{P]})F_{2R]}{}^T\nm
&-&(5F_4{}^{NO}{}_{|T|,R]}+6F_4{}^N{}_{R]T,}{}^O
-F_4{}^{NO}{}_{R],T}))F_2{}^{P]T}\nm
&+&\frac{2}{3(4+n)}\h^{[M}_{\phantom{[}Q}\h^N_R(
F_4{}^{OP]}{}_{U,V}-F_{4UV}{}^{O,P]})F_2{}^{QR})\nm
&+&\Lambda((F_4{}^{MNOP}{}_{V,[Q}
+F_4{}^{[MNO}{}_{[Q|V|,}{}^{P]})F_{2R]}{}^V\nm 
&+&(F_4{}^{[MN}{}_{QRV,}{}^O
-F_4{}^{[MNO}{}_{[Q|V|,R]})F_2{}^{P]V}\nm
&-&\h^{[M}_{[Q}(F_4{}^{NO}{}_{R]UV,}{}^{P]}
+F_4{}^{NOP]}{}_{UV,R})F_2{}^{UV})\nm 
&+&\Sigma(F_3{}^{[MNO}{}_TF_3{}^{P]}{}_{QR}{}^T
-F_3{}^{[MN}{}_{TQ}F_3{}^{OP]T}{}_R\nm
&+&\frac{10}{(4+n)}\h^{[M}_{[Q}F_3{}^{NO}{}_{TS}
F_3{}^{P]}{}_{R]}{}^{TS}\nm
&+&\frac{10}{(4+n)(5+n)}\h^{[M}_{\phantom{[}Q}\h^{N}_R
F_3{}^{O}{}_{TSU}F_3{}^{P]TSU} )
\eea 

Constraints

\bea
-\Gamma-\frac{4}{3}\Sigma -\frac{3}{2}\Delta + \frac{3}{10}\frac{(10+n)}{4+n}\Lambda &=&0\nm
\Gamma+\Sigma +\frac{1}{2}\Delta - \frac{3}{10}\frac{(10+2n)}{4+n}\Lambda &=&0
\eea



\begin{thebibliography}{99}

\bibitem{Cremmer:1979up}
  E.~Cremmer and B.~Julia,
  ``The SO(8) Supergravity,''
  Nucl.\ Phys.\  B {\bf 159} (1979) 141.
  
  
\bibitem{Marcus:1983hb}
  N.~Marcus and J.~H.~Schwarz,
  ``Three-Dimensional Supergravity Theories,''
  Nucl.\ Phys.\  B {\bf 228} (1983) 145.
  
\bibitem{Greitz:2011vh}
  J.~Greitz and P.~S.~Howe,
  ``Maximal supergravity in three dimensions: supergeometry and differential forms,''
  JHEP {\bf 1107} (2011) 071
  [arXiv:1103.2730 [hep-th]].
  
\bibitem{HenryLabordere:2002dk}
  P.~Henry-Labordere, B.~Julia and L.~Paulot,
  ``Borcherds symmetries in M theory,''
  JHEP {\bf 0204} (2002) 049
  [hep-th/0203070].

\bibitem{HenryLabordere:2002xh}
  P.~Henry-Labordere, B.~Julia and L.~Paulot,
  ``Real Borcherds superalgebras and M-theory,''
  JHEP {\bf 0304} (2003) 060
  [arXiv:hep-th/0212346].
  
\bibitem{Julia:1997cy}
  B.~L.~Julia,
  ``Dualities in the classical supergravity limits: Dualisations,  dualities
  and a detour via 4k+2 dimensions,''
  arXiv:hep-th/9805083.
  
 
  
\bibitem{West:2001as}
  P.~C.~West,
  ``E(11) and M theory,''
  Class.\ Quant.\ Grav.\  {\bf 18} (2001) 4443
  [arXiv:hep-th/0104081].
  
\bibitem{Riccioni:2007au}
  F.~Riccioni and P.~C.~West,
  ``The E(11) origin of all maximal supergravities,''
  JHEP {\bf 0707} (2007) 063
  [arXiv:0705.0752 [hep-th]].
  
\bibitem{de Wit:1992up}
  B.~de Wit, A.~K.~Tollsten and H.~Nicolai,
  ``Locally supersymmetric D = 3 nonlinear sigma models,''
  Nucl.\ Phys.\ B {\bf 392} (1993) 3
  [hep-th/9208074].
  
\bibitem{Bergshoeff:2007vb}
  E.~A.~Bergshoeff, J.~Gomis, T.~A.~Nutma and D.~Roest,
  ``Kac-Moody Spectrum of (Half-)Maximal Supergravities,''
  JHEP {\bf 0802} (2008) 069
  [arXiv:0711.2035 [hep-th]].
  
  
\bibitem{Nicolai:2001ac}
  H.~Nicolai and H.~Samtleben,
  ``N=8 matter coupled AdS(3) supergravities,''
  Phys.\ Lett.\ B {\bf 514} (2001) 165
  [hep-th/0106153].
  
\bibitem{deWit:2003ja}
  B.~de Wit, I.~Herger and H.~Samtleben,
  ``Gauged locally supersymmetric D = 3 nonlinear sigma models,''
  Nucl.\ Phys.\ B {\bf 671} (2003) 175
  [hep-th/0307006].
  
     
  
\bibitem{deWit:2008gc}
  B.~de Wit and H.~Samtleben,
  ``The end of the p-form hierarchy,''
  JHEP {\bf 0808} (2008) 015
  [arXiv:0805.4767 [hep-th]].
  
\bibitem{Greitz:2011da}
  J.~Greitz and P.~S.~Howe,
  ``Maximal supergravity in D=10: Forms, Borcherds algebras and superspace cohomology,''
  JHEP {\bf 1108} (2011) 146
  [arXiv:1103.5053 [hep-th]].
  
\bibitem{Slansky:1991dx}
  R.~Slansky,
  ``An Algebraic Role For Energy And Number Operators For Multiparticle
  States,''
  Nucl.\ Phys.\  B {\bf 389} (1993) 349.
  
 
  
  
   
\bibitem{Howe:1995zm}
  P.~S.~Howe, J.~M.~Izquierdo, G.~Papadopoulos and P.~K.~Townsend,
  ``New supergravities with central charges and Killing spinors in
  (2+1)-dimensions,''
  Nucl.\ Phys.\  B {\bf 467} (1996) 183
  [arXiv:hep-th/9505032].
  
\bibitem{Howe:2004ib}
  P.~S.~Howe and E.~Sezgin,
  ``The Supermembrane revisited,''
  Class.\ Quant.\ Grav.\  {\bf 22} (2005) 2167
  [arXiv:hep-th/0412245].
  
\bibitem{Kuzenko:2011xg}
  S.~M.~Kuzenko, U.~Lindstrom and G.~Tartaglino-Mazzucchelli,
  ``Off-shell supergravity-matter couplings in three dimensions,''
  JHEP {\bf 1103} (2011) 120
  [arXiv:1101.4013 [hep-th]].
   
  
\bibitem{Cederwall:2011pu}
  M.~Cederwall, U.~Gran and B.~E.~W.~Nilsson,
  ``D=3, N=8 conformal supergravity and the Dragon window,''
  arXiv:1103.4530 [hep-th].
  
\bibitem{Dragon:1978nf}
  N.~Dragon,
  ``Torsion And Curvature In Extended Supergravity,''
  Z.\ Phys.\  C {\bf 2} (1979) 29.
 
 
 
\bibitem{Weidner:2006rp}
  M.~Weidner,
  ``Gauged supergravities in various spacetime dimensions,''
  Fortsch.\ Phys.\  {\bf 55} (2007) 843
  [hep-th/0702084].
  
\bibitem{Nicolai:2000sc}
  H.~Nicolai and H.~Samtleben,
  ``Maximal gauged supergravity in three dimensions,''
  Phys.\ Rev.\ Lett.\  {\bf 86} (2001) 1686
  [arXiv:hep-th/0010076].
  
\bibitem{Nicolai:2001sv}
  H.~Nicolai and H.~Samtleben,
  ``Compact and noncompact gauged maximal supergravities in three
  dimensions,''
  JHEP {\bf 0104} (2001) 022
  [arXiv:hep-th/0103032].
  
\bibitem{de Wit:1982ig}
  B.~de Wit and H.~Nicolai,
  ``N=8 Supergravity,''
  Nucl.\ Phys.\  B {\bf 208} (1982) 323.

  
\bibitem{Howe:1981tp}
  P.~S.~Howe and H.~Nicolai,
  ``Gauging N=8 Supergravity In Superspace,''
  Phys.\ Lett.\  B {\bf 109} (1982) 269.
      

\bibitem{Cremmer:1997ct}
  E.~Cremmer, B.~Julia, H.~Lu and C.~N.~Pope,
  ``Dualisation of dualities. I,''
  Nucl.\ Phys.\  B {\bf 523} (1998) 73
  [arXiv:hep-th/9710119].
  
\bibitem{Cremmer:1998px}
  E.~Cremmer, B.~Julia, H.~Lu and C.~N.~Pope,
  ``Dualisation of dualities. II: Twisted self-duality of doubled fields  and
  superdualities,''
  Nucl.\ Phys.\  B {\bf 535} (1998) 242
  [arXiv:hep-th/9806106].

\bibitem{Iqbal:2000}
  A.~Iqbal, A.~Neitzke and C.~Vafa
  ``A Mysterious Duality,''
  Adv.\ Theor.\ Math.\ Phys.  {\bf 5} (2002) 769
  [	arXiv:hep-th/0111068v2].


\bibitem{Henneaux:2010ys}
  M.~Henneaux, B.~L.~Julia and J.~Levie,
  ``$E_{11}$, Borcherds algebras and maximal supergravity,''
  arXiv:1007.5241 [hep-th].

\bibitem{Palmkvist:2011vz}
  J.~Palmkvist,
  ``Tensor hierarchies, Borcherds algebras and E11,''
  JHEP {\bf 1202} (2012) 066
  [arXiv:1110.4892 [hep-th]].

\bibitem{Palmkvist:2012}
  J.~Palmkvist,
  ``Borcherds and Kac-Moody extensions of simple finite-dimensional Lie algebras,''
  [arXiv:1203.5107 [hep-th]]. 


\bibitem{Nutma}
  T.~Nutma
  ``SimpLie-a simple program for Lie algebras,'' http://code.google.com/p/simplie/
   

\bibitem{Bergshoeff:2008}
  E.~Bergshoeff, J.~Gomis, T.~Nutma and D.~Roest,
  ``Kac-Moody Spectrum of (Half-)Maximal Supergravities,''
  JHEP {\bf 0802} (2008) 069
  [arXiv:hep-th/0711:2035].
  
\bibitem{Damour:2002}
  T.~Damour, M.~Henneaux and H.~Nicolai
  ``E10 and a "small tension expansion" of M Theory''
  Phys.\ Rev.\ Lett. {\bf 89} (2002) 221601
  [arXiv:hep-th/0207267v1].


\bibitem{Brink:1979nt}
  L.~Brink and P.~S.~Howe,
  ``The N=8 Supergravity In Superspace,''
  Phys.\ Lett.\  B {\bf 88} (1979) 268.
  
  
\bibitem{Howe:1981gz}
  P.~S.~Howe,
  ``Supergravity In Superspace,''
  Nucl.\ Phys.\  B {\bf 199} (1982) 309.
  
  
\bibitem{deWit:1983gs}
  B.~de Wit and H.~Nicolai,
  ``The Parallelizing S(7) Torsion In Gauged N=8 Supergravity,''
  Nucl.\ Phys.\  B {\bf 231} (1984) 506.
  
\bibitem{Bonora:1986ix}
  L.~Bonora, P.~Pasti and M.~Tonin,
  ``Superspace Formulation Of 10-D Sugra+Sym Theory A La Green-Schwarz,''
  Phys.\ Lett.\  B {\bf 188} (1987) 335.
  
 
  
\bibitem{Howe:1991mf}
  P.~S.~Howe,
  ``Pure Spinors Lines In Superspace And Ten-Dimensional Supersymmetric
  Theories,''
  Phys.\ Lett.\  B {\bf 258} (1991) 141
  [Addendum-ibid.\  B {\bf 259} (1991) 511].

\bibitem{Howe:1991bx}
  P.~S.~Howe,
  ``Pure Spinors, Function Superspaces And Supergravity Theories In
  Ten-Dimensions And Eleven-Dimensions,''
  Phys.\ Lett.\  B {\bf 273} (1991) 90.
  
\bibitem{Berkovits:2002zk}
  N.~Berkovits,
  ``ICTP lectures on covariant quantization of the superstring,''
  arXiv:hep-th/0209059.
 


\bibitem{Berkovits:2008qw}
  N.~Berkovits and P.~S.~Howe,
  ``The cohomology of superspace, pure spinors and invariant integrals,''
  arXiv:0803.3024 [hep-th].
  
\bibitem{Brandt:2010fa}
  F.~Brandt,
  ``Supersymmetry Algebra Cohomology: II. Primitive Elements in 2 and 3 Dimensions,''
  J.\ Math.\ Phys.\  {\bf 51} (2010) 112303
  [arXiv:1004.2978 [hep-th]].
  
\bibitem{Movshev:2011pr}
  M.~V.~Movshev, A.~Schwarz and R.~Xu,
  ``Homology of Lie algebra of supersymmetries and of super Poincare Lie algebra,''
  Nucl.\ Phys.\ B {\bf 854} (2012) 483
  [arXiv:1106.0335 [hep-th]].
  
  
  
    
      
    
 \end{thebibliography}
\end{document}